%% file: surf_op_ZM_a.tex
\pdfoutput=1
\documentclass[11pt,a4paper]{article}
\usepackage{cite}
\usepackage{graphicx}
\usepackage{amssymb}
\usepackage{amsmath}
\usepackage{amsfonts}
\usepackage{dsfont}
\usepackage{mathtools}
\usepackage{slashed}
\usepackage{rotating}
\usepackage{bbold,amsfonts}

\usepackage[utf8]{inputenc}
\usepackage{bm}
\usepackage{xcolor}
\usepackage{float}
\usepackage{braket}

\usepackage[height=8.8in,width=6.45in]{geometry}
\usepackage[font=small,labelfont=bf]{caption}
\usepackage[hidelinks]{hyperref}
\numberwithin{equation}{section}

\newcommand{\beq}{\begin{equation}}
\newcommand{\eeq}{\end{equation}}

\newcommand{\ii}{\mathrm{i}}
\newcommand{\overbar}[1]{\mkern 1.5mu\overline{\mkern-1.5mu#1\mkern-1.5mu}\mkern 1.5mu}
\makeatletter
\newcommand*{\letterdef@}{}
\newcommand*{\letterdef}[3]{%
	\def\letterdef@##1{\expandafter\newcommand\csname #1\endcsname{#2{##1}}}%
	\@tfor\@tempa :=#3\do{\expandafter\letterdef@\expandafter{\@tempa}}}
\makeatother
\letterdef{c#1} {\mathcal}{ABCDEFGHIJKLMNOPQRSTUVWXYZ} 
\letterdef{rm#1}{\mathrm} {dDeimM} 

\begin{document}
\begin{titlepage}
\vbox{
    \halign{#\hfil         \cr
           } 
      }  
\vspace*{15mm}
\begin{center}
{\LARGE \bf 
Surface Defects from Fractional Branes - II
}

\vspace*{15mm}

{\Large S.~K.~Ashok$\,{}^{a}$,  M.~Bill\`o$\,{}^{b,c}$, M.~Frau$\,{}^{b,c}$,
	A.~Lerda$\,{}^{d,c}$ and S.~Mahato$\,{}^{a}$}
\vspace*{8mm}

${}^a$ Institute of Mathematical Sciences, \\
          Homi Bhabha National Institute (HBNI),\\
		 IV Cross Road, C.I.T. Campus, \\
		  Taramani, Chennai, India 600113
		  \vskip 0.5cm
			
${}^b$ Universit\`a di Torino, Dipartimento di Fisica,\\
		   Via P. Giuria 1, I-10125 Torino, Italy
		   \vskip 0.5cm

${}^c$ I.N.F.N. - sezione di Torino,\\
		   Via P. Giuria 1, I-10125 Torino, Italy 
		   \vskip 0.5cm		

${}^d$  Universit\`a del Piemonte Orientale,\\
			Dipartimento di Scienze e Innovazione Tecnologica\\
			Viale T. Michel 11, I-15121 Alessandria, Italy			
	
\vskip 0.8cm
	{\small
		E-mail:
		\texttt{sashok,sujoymahato@imsc.res.in; billo,frau,lerda@to.infn.it}
	}
\vspace*{0.8cm}
\end{center}

\begin{abstract}

A generic half-BPS surface defect of ${\mathcal N}=4$ supersymmetric U$(N)$ Yang-Mills theory is described by a partition of $N = n_1 + \ldots + n_M$ and a set of $4M$ continuous parameters. We show that such a defect can be realized by $n_I$ stacks of fractional D3-branes in Type II B string theory on a $\mathbb{Z}_M$ orbifold background in which the brane world-volume is partially extended along the orbifold directions. In this set up we show that the $4M$ continuous parameters correspond to constant background values of certain twisted closed string scalars of the orbifold. These results extend and generalize what we have presented for the simple defects in a previous paper.

\end{abstract}
\end{titlepage}
\setcounter{tocdepth}{2}
\tableofcontents
\vspace{1cm}
\begingroup
\allowdisplaybreaks

\section{Introduction}

In this article, which is a direct extension of the companion paper \cite{Ashok:2020ekv}, we discuss
how to embed into string theory a generic Gukov-Witten (GW) 
surface defect \cite{Gukov:2006jk,Gukov:2008sn}  of the ${\mathcal N}=4$  
supersymmetric U$(N)$ Yang-Mills theory. 
We do this by analyzing the massless fields on the world-volume 
of fractional D3-branes in Type II B string theory on an orbifold background, 
following the proposal of Kanno-Tachikawa (KT)\cite{Kanno:2011fw}. 
Here, we shall only provide a brief introduction and refer the reader to the introductory 
sections of the companion paper \cite{Ashok:2020ekv} for a more detailed account and for
a discussion of the relation of our approach to others already 
present in the literature as well as for the relevant references.
 
We consider Type II B string theory on the following orbifold space-time
\begin{equation}
{\mathbb C}_{(1)} \times \frac{{\mathbb C}_{(2)} \times {\mathbb C}_{(3)} }{\mathbb{Z}_M} \times {\mathbb C}_{(4)} \times {\mathbb C}_{(5)}
\label{orbifold}
\end{equation}
with constant vacuum expectation values turned on for particular twisted scalar fields in the 
Neveu-Schwarz/Neveu-Schwarz (NS/NS) and Ramond/Ramond (R/R) sectors. In this background
we engineer a 4$d$ gauge theory by introducing stacks of fractional D3-branes that extend 
along the first two complex planes $\mathbb{C}_{(1)}$ and ${\mathbb C}_{(2)}$. 
In this combined orbifold/D-brane set-up, which we refer to as the KT configuration 
\cite{Kanno:2011fw}, we compute the profile in configuration space of the massless open strings 
by means of open/closed world-sheet correlators, and show that these exactly reproduce the 
singular profiles that characterize the GW surface defect in the ${\mathcal N}=4$ gauge theory \cite{Gukov:2006jk}. 
In this way we are therefore able to provide an explicit identification of the continuous parameters
of the GW solution with the vacuum expectation values of the twisted scalars.

In \cite{Ashok:2020ekv} we already worked out this identification for the simple surface defects 
that correspond to the ${\mathbb Z}_2$ orbifold. In this paper, we extend our
analysis to the ${\mathbb Z}_M$ orbifolds for $M>2$ which can describe the most 
general defect corresponding to the breaking of the $U(N)$ gauge 
group to the Levi subgroup $\mathrm{U}(n_0)\times\ldots\times\mathrm{U}(n_{M-1})$ 
with $\sum_In_I=N$.

While the basic conceptual issues in realizing such a surface defect using fractional D3-branes 
remain the same for all $M$, the main difference with respect to \cite{Ashok:2020ekv} lies 
in the treatment of the closed string background. 
For $M = 2$ the massless fields of the NS/NS and R/R twisted sectors 
correspond to degenerate ground states and their vertex operators are realized using spin fields
\cite{Friedan:1985ge}. This is no longer the case for $M > 2$ and, in fact, the massless fields 
of the NS/NS sector arise from excited
states created by the oscillators of the fermionic string coordinates. Furthermore, pairs of twisted 
sectors are related by complex conjugation and this turns out to play an important role in the
identification of the closed string background with the real parameters in the GW profiles.

In the $\mathbb{Z}_M$ orbifold, there are $(M-1)$ twisted sectors. One could treat all of them at once using the bosonization formalism \cite{Friedan:1985ge,Kostelecky:1986xg}, but in order to
keep track of all the relative phases it would be necessary to introduce the so-called cocycle factors.
Since dealing with these cocycle factors is quite involved, and since the relative phases are crucial 
to obtain the correct identification of the continuous parameters of the GW surface defects,
we adopt an explicit fermionic approach and use the bosonization formalism only where no 
phase ambiguities arise. The advantage of this method is that the relative phases among 
the contributions from different sectors are easily tracked and fixed by the fermionic statistics. Moreover, in this fermionic  approach we can describe the fractional D3-branes using boundary states (for a review see for example \cite{DiVecchia:1999mal,DiVecchia:1999fje}). Even though the KT brane configuration
has not been explicitly considered so far from the boundary state point of view, we can exploit 
many of the results that already exist in the literature 
\cite{Douglas:1996sw,Billo:2000yb,Billo:2001vg,Bertolini:2001gq} and generalize 
them to the present case, in which the fractional D3-branes partially extend along the orbifold. The price we 
have to pay for using this fermionic approach is that we have to distinguish between the 
twisted sectors and treat separately those whose twist parameter is smaller or bigger than 
$\frac{1}{2}$.

The open string sector, instead, is similar to that of the $M=2$ case. We recall that for the KT configuration, the fractional D3-branes have the same field content as the regular D3-branes, 
since in this case the orbifold does not project away any of the open string excitations, unlike the case when the branes are entirely transverse to the orbifolded space. Indeed, on the world-volume of the fractional 
D3-branes we find a gauge vector and three complex massless scalars plus their fermionic partners. 
However, the corresponding vertex operators are linear combinations that behave covariantly under the action of the orbifold. When $M>2$, these combinations are slightly more involved than for $M=2$ and are written in terms of generalized plane-waves. 

Once the vertex operators for the massless open and closed string states are derived, the discussion proceeds along the same lines as in the $M=2$ case, but with the important technical differences and peculiarities that we have just mentioned. 

This paper is structured as follows.
In Section~\ref{closedspectrum} we provide a detailed description of
the twisted closed string spectrum of Type II B strings on the orbifold (\ref{orbifold}). We then 
proceed in Section~\ref{Dbranes} to introduce the fractional D3-branes of the $\mathbb{Z}_M$
orbifold and study two different aspects. Firstly, from the closed string point of view we write the
boundary states and use them to derive the reflection rules that relate the left- and right-moving modes of the twisted closed strings in all sectors. Secondly, we derive the $\mathbb{Z}_M$-invariant 
vertex operators describing the massless open string excitations that live on the D3-brane 
world-volume. This turns out to be non-trivial given that the D3-brane extends partially along the orbifolded space. In Section~\ref{diskcorr} we calculate mixed correlators among the 
open and closed string excitations and use them to derive in Section~\ref{profilesfromFT}
the singular profiles of the gauge fields near the location of the surface defect.
The field profiles we obtain precisely match those of the GW defect once the background
values of the twisted closed string field are identified with the continuous parameters of the surface defect. 

Our analysis provides an explicit realization of the monodromy defects of the $\mathcal{N}=4$
super Yang-Mills theory using perturbative string theory methods. As we discuss in the
concluding Section~\ref{sec:concl-m} we believe that this stringy realization may prove to be useful
in further investigations of surface defects and their properties, and it may even offer an alternative approach to the study of extended objects in ordinary field theory through their embedding into string theory.

\section{Twisted closed strings in the $\mathbb{Z}_M$ orbifold}
\label{closedspectrum}

We consider Type II~B string theory on the orbifold (\ref{orbifold}).
The $i$-th complex plane ${\mathbb C}_{(i)}$ is parametrized by
\begin{equation}
z_i = \frac{x_{2i-1} +\ii\, x_{2i}}{\sqrt{2}} \qquad\text{and}\qquad \bar z_i 
= \frac{x_{2i-1} -\ii\, x_{2i}}{\sqrt{2}}
\label{zi}
\end{equation}
where $x_{\mu}$ are the ten real coordinates of space-time. The orbifold group $\mathbb{Z}_M$
is generated by an element $g$ such that $g^M=1$, with the following action on $z_2$ and $z_3$:
\begin{equation}
g\,:~ (z_2\,, \,z_3) \longrightarrow (\omega\, z_2\,,\, \omega^{-1} z_3) 
\label{gonz23}
\end{equation}
where
\begin{equation}
\omega = \rme^{\frac{2\pi\ii}{M}}~.
\label{omega}
\end{equation}
The action of $g$ on $\bar z_2$ and $\bar z_3$ follows from complex conjugation. This breaks
the $\mathrm{SO}(4)\simeq \mathrm{SU}(2)_+\times\mathrm{SU}(2)_-$ isometry group of
$\mathbb{C}_{(2)}\times\mathbb{C}_{(3)}$ to SU(2)$_+$.

To describe the closed strings propagating on this orbifold, we use the complex notation and denote 
the bosonic string coordinates by $\left\{ Z^i(z), \overbar{Z}^i(z)\right \}$ for the left-movers and 
$\big\{\widetilde{Z^i}(\bar z), \widetilde{\overbar{Z}^i}(\bar z)\big\}$ for the 
right-movers, with $z$
and $\bar{z}$ parametrizing the closed string world-sheet.
Similarly, we denote the fermionic string coordinates by $\big\{\Psi^i(z), \overbar{\Psi}^i(z)\big\}$ 
for the left-movers and $\big\{\widetilde{\Psi^i}(\bar z), \widetilde{\overbar{\Psi}^i}(\bar z)\big\}$ 
for the right-movers. 

\subsection{Twisted sectors}
In the $\mathbb{Z}_M$ orbifold theory, there are $(M-1)$ twisted sectors labeled by the
index $\widehat{a}=1,\ldots,M-1$. If $M$ is odd, 
we can divide the twisted sectors in two sets, each one containing $\frac{M-1}{2}$ elements.
The sectors of the first set are labeled by $\widehat{a}=a=1,\cdots,\frac{M-1}{2}$ and
are characterized by a twist parameter 
\begin{equation}
\nu_a = \frac{a}{M}<\frac{1}{2}~.
\label{twistless12}
\end{equation}
The sectors of the second set have, instead,  a twist parameter
\begin{equation}
1-\nu_a =\frac{M-a}{M}>\frac{1}{2}~
\label{twistmore12}
\end{equation}
and are labeled by $\widehat{a}=(M-a)$.
If $M$ is even, in addition there is an extra sector with twist parameter 
$\frac{1}{2}$, which has to be treated separately. For most of the discussion we will assume that 
$M$ is odd and briefly comment on the special case with twist $\frac{1}{2}$, occurring when $M$ is even, only at the very end, since this case has already been discussed in detail in the companion paper \cite{Ashok:2020ekv}.

In the sectors with label $a$ and twist parameter as in (\ref{twistless12}), the left-movers of the bosonic and fermionic string coordinates satisfy the following monodromy properties on the world-sheet:
\begin{subequations}
\label{monodleft}
\begin{align}
Z^2(\rme^{2\pi\ii}\, z)&= \rme^{2\pi\ii\nu_a}\, Z^2(z)~, 
\qquad Z^3(\rme^{2\pi\ii}\, z)= \rme^{-2\pi\ii\nu_a}\, Z^3(z)~,\\[1mm]
\Psi^2(\rme^{2\pi\ii}\, z)&= \pm \,\rme^{2\pi\ii\nu_a}\, \Psi^2(z)~, 
\quad\,\Psi^3(\rme^{2\pi\ii}\, z)= \pm \,\rme^{-2\pi\ii\nu_a}\, \Psi^3(z)~,
\label{psi2psi3mono}
\end{align}%
\end{subequations}
where the $+ (-)$ sign refers to the NS (R) sector. 
The analogous relations for $\overbar{Z}^i$ and $\overbar{\Psi}^i$ can be obtained by 
complex conjugation.
On the other hand, the right-movers satisfy the monodromy properties:
\begin{subequations}
\label{monodright}
\begin{align}
\widetilde{Z^2}(\rme^{2\pi\ii}\, \bar z) &= \rme^{-2\pi\ii\nu_a}\, \widetilde{Z^2}(\bar z)~, 
\qquad \widetilde{Z^3}(\rme^{2\pi\ii}\, \bar z)= \rme^{2\pi\ii\nu_a}\, \widetilde{Z^3}(\bar z)~,\\[1mm]
\widetilde{\Psi^2}(\rme^{2\pi\ii}\, \bar z) &= \pm\, \rme^{-2\pi\ii\nu_a}\, \widetilde{\Psi^2}(\bar z)~, 
\quad\, \widetilde{\Psi^3}(\rme^{2\pi\ii}\, \bar z)= \pm\, \rme^{2\pi\ii\nu_a}\, 
\widetilde{\Psi^3}(\bar z)~.
\label{psi2psi3rightmono}
\end{align}%
\end{subequations}
Again, the relations for $\widetilde{\overbar{Z}^i}$ and $\widetilde{\overbar{\Psi}^i}$ are 
obtained by complex conjugation.

For the sectors with label $(M-a)$ and twist parameter as in (\ref{twistmore12}), similar monodromy relations hold for the world-sheet fields but with $\nu_a$ everywhere replaced by $(1-\nu_a)$. 

\subsection{Twisted NS sectors }
\label{twistNS}
We now turn to a discussion of the spectrum of massless string states in the various 
twisted sectors, focusing mainly on the fermionic fields in the complex directions 2 and 3.
In the fermionic formalism, when the NS boundary conditions are imposed, we have to treat 
separately the sectors with twist 
parameter smaller than $\frac{1}{2}$ and those with twist parameter bigger than $\frac{1}{2}$.

\subsubsection{Sectors with twist parameter $\nu_a<\frac{1}{2}$}
\label{secn:NSNSa}
In this case the monodromy properties (\ref{psi2psi3mono}) and their complex conjugate 
lead to the following mode expansions for the left-moving fermionic fields
(see for example \cite{Bertolini:2005qh} and references therein):
\begin{equation}
\begin{aligned}
\Psi^2(z)&=\sum_{r={1}/{2}}^\infty\Big(\overbar{\Psi}^{2}_{r-\nu_a}\,z^{-r+\nu_a-\frac{1}{2}}
+{\Psi}^{2}_{-r-\nu_a}\,z^{r+\nu_a-\frac{1}{2}}
\Big)~,\\
\overbar{\Psi}^{2}(z)&=\sum_{r={1}/{2}}^\infty\Big({\Psi}^{2}_{r+\nu_a}\,z^{-r-\nu_a-\frac{1}{2}}
+\overbar{\Psi}^{2}_{-r+\nu_a}\,z^{r-\nu_a-\frac{1}{2}}
\Big)~,
\end{aligned}
\label{Psi2a}
\end{equation}
and
\begin{equation}
\begin{aligned}
\Psi^3(z)&=\sum_{r={1}/{2}}^\infty\Big(\overbar{\Psi}^{3}_{r+\nu_a}\,z^{-r-\nu_a-\frac{1}{2}}
+{\Psi}^{3}_{-r+\nu_a}\,z^{r-\nu_a-\frac{1}{2}}
\Big)~,\\
\overbar{\Psi}^{3}(z)&=\sum_{r={1}/{2}}^\infty\Big({\Psi}^{3}_{r-\nu_a}\,z^{-r+\nu_a-\frac{1}{2}}
+\overbar{\Psi}^{3}_{-r-\nu_a}\,z^{r+\nu_a-\frac{1}{2}}
\Big)~.
\end{aligned}
\label{Psi3a}
\end{equation}
The oscillators ${\Psi}^{2}_{-r-\nu_a}$, $\overbar{\Psi}^{2}_{-r+\nu_a}$, ${\Psi}^{3}_{-r+\nu_a}$ and
$\overbar{\Psi}^{3}_{-r-\nu_a}$ are creation modes acting on the twisted vacuum of the $a$-th
sector which we denote by $|\Omega_{a}\rangle$. Such a state is defined by
\begin{equation}
|\Omega_{a}\rangle =\lim_{z\to 0}\,\sigma_{a}(z)\,s_{a}(z)\,|0\rangle
\end{equation}
where $|0\rangle$ is the Fock vacuum and  $\sigma_{a}(z)$ and $s_{a}(z)$ are, respectively, the bosonic and fermionic twist fields \cite{Dixon:1986qv}. More precisely, these twist fields take the form
\begin{equation}
\begin{aligned}
\sigma_{a}(z)& =\sigma_{\nu_a}^2(z)\,\sigma_{1-\nu_a}^3(z)\quad\text{and}\quad
s_{a}(z) = s_{\nu_a}^2(z)\,s_{-\nu_a}^3(z)~,
\end{aligned}
\label{sigmas}
\end{equation}
where the superscripts refer to the complex directions where the twist takes place, and the subscripts
indicate the twist parameters.
The bosonic twist field $\sigma_{a}(z)$ 
is a conformal field of weight $\nu_a(1-\nu_a)$ while the fermionic
twist field $s_{a}(w)$ is a conformal field of weight $\nu_a^2$. Therefore, the total conformal weight of the operator associated to the twisted ground state is $\nu_a$. This means that $|\Omega_{a}\rangle$ is massive with a mass $m$ given by
\begin{equation}
m^2=\nu_a-\frac{1}{2}<0~.
\label{massground}
\end{equation}
This tachyonic state is removed by the GSO projection.

The first set of physical states one finds in the GSO projected spectrum are those obtained 
by acting with one fermionic creation mode with index $r=\frac{1}{2}$ on the twisted vacuum. 
In particular, the oscillators ${\Psi}^{3}_{-\frac{1}{2}+\nu_a}$ and $\overbar{\Psi}^{2}_{-\frac{1}{2}+\nu_a}$ increase the energy
by $(\frac{1}{2}-\nu_a)$ and thus, when acting on the twisted vacuum, they create two
massless states\,%
\footnote{The oscillators ${\Psi}^{2}_{-\frac{1}{2}-\nu_a}$ and $\overbar{\Psi}^{3}_{-\frac{1}{2}-\nu_a}$,
instead, carry an energy $(\frac{1}{2}+\nu_a)$ and, upon acting on the twisted vacuum, they create
massive states with $m^2=2\nu_a$.}. The vertex operators corresponding to these massless excitations, in the $(-1)$-superghost picture and at zero momentum\,%
\footnote{The reason to write the vertex operators at zero momentum is because, as in 
\cite{Ashok:2020ekv}, ultimately we will be interested in describing a constant twisted closed 
string background to account for the continuous parameters of the GW surface defects.}, are:
\begin{equation}
\begin{aligned}
\mathcal{V}^1_a(z)&=\sigma_{a}(z):\!{\Psi}^{3}(w)\,s_{a}(z)\!:\rme^{-\phi(z)}~,\\[1mm]
\mathcal{V}^2_a(z)&=\sigma_{a}(z):\!\overbar{\Psi}^{\,2}(w)\,s_{a}(z)\!:\rme^{-\phi(z)}~.
\end{aligned}
\label{vertex1}
\end{equation}
Here $\phi(z)$ is the bosonic field appearing in the bosonization formulas of the superghosts
\cite{Friedan:1985ge} and, as usual, the symbol :~: denotes the normal ordering.
The vertex operators (\ref{vertex1})  are conformal fields of weight 1 and we collectively 
denote them as $\mathcal{V}^\alpha_a(z)$ with $\alpha=1,2$.
As explained in Appendix~\ref{SO(4)}, they form a doublet transforming as a spinor of SU(2)$_+$.

In the right-moving part, the monodromy properties (\ref{psi2psi3rightmono}) lead to the
following mode expansions for the fermionic fields
\begin{equation}
\begin{aligned}
\widetilde{\Psi^2}(\bar{z})&=\sum_{r={1}/{2}}^\infty\Big(\widetilde{\overbar{\Psi}^2}_{\!r+\nu_a}\,\bar{z}^{-r-\nu_a-\frac{1}{2}}
+\widetilde{\Psi^2}_{\!-r+\nu_a}\,\bar{z}^{r-\nu_a-\frac{1}{2}}
\Big)~,\\
\widetilde{\overbar{\Psi}^2}(\bar{z})&=\sum_{r={1}/{2}}^\infty\Big(\widetilde{\Psi^2}_{\!r
-\nu_a}\,\bar{z}^{-r+\nu_a-\frac{1}{2}}
+\widetilde{\overbar{\Psi}^2}_{\!-r-\nu_a}\,\bar{z}^{r+\nu_a-\frac{1}{2}}
\Big)~,
\end{aligned}
\end{equation}
and
\begin{equation}
\begin{aligned}
\widetilde{\Psi^3}(\bar{z})&=\sum_{r={1}/{2}}^\infty\Big(\widetilde{\overbar{\Psi}^3}_{\!r-\nu_a}\,
\bar{z}^{-r+\nu_a-\frac{1}{2}}
+\widetilde{\Psi^3}_{\!-r-\nu_a}\,\bar{z}^{r+\nu_a-\frac{1}{2}}
\Big)~,\\
\widetilde{\overbar{\Psi}^3}(\bar{z})&=\sum_{r={1}/{2}}^\infty\Big(\widetilde{\Psi^3}_{\!r
+\nu_a}\,\bar{z}^{-r-\nu_a-\frac{1}{2}}
+\widetilde{\overbar{\Psi}^3}_{\!-r+\nu_a}\,\bar{z}^{r-\nu_a-\frac{1}{2}}
\Big)~.
\end{aligned}
\end{equation}
The oscillators $\widetilde{\Psi^2}_{\!-r+\nu_a}$, $\widetilde{\overbar{\Psi}^2}_{\!-r-\nu_a}$, 
$\widetilde{\Psi^3}_{\!-r-\nu_a}$ and $\widetilde{\overbar{\Psi}^3}_{\!-r+\nu_a}$ are creation modes
acting on the twisted vacuum of the right sector which we denote by $|\widetilde{\Omega}_{a}\rangle$. This is defined by 
\begin{equation}
|\widetilde{\Omega}_{a}\rangle =\lim_{\bar{z}\to 0}\,\widetilde{\sigma}_{a}(\bar{z})\,
\widetilde{s}_{a}(\bar{z})\,|\widetilde{0}\rangle
\end{equation}
where $|\widetilde{0}\rangle$ is the Fock vacuum of this sector and
\begin{equation}
\begin{aligned}
\widetilde{\sigma}_{a}(\bar{z})& =\,\widetilde{\sigma}_{1-\nu_a}^{\,2}(\bar{z})\,
\widetilde{\sigma}_{\nu_a}^{\,3}(\bar{z})\quad\text{and}\quad 
\widetilde{s}_{a}(\bar{z})&=\,\widetilde{s}_{-\nu_a}^{\,2}(\bar{z})\,\widetilde{s}_{\nu_a}^{\,3}(\bar{z})~.
\end{aligned}
\label{sigmasr}
\end{equation}
The bosonic twist field $\widetilde{\sigma}_{a}(\bar{z})$ is a conformal field of weight 
$(1-\nu_a)\nu_a$ while the fermionic
twist field $\widetilde{s}_{a}(\bar{z})$ is a conformal field of weight $\nu_a^2$, so that the total
conformal weight of the 
operator associated to $|\widetilde{\Omega}_{a}\rangle$ is $\nu_a$.
The right-moving ground state is then tachyonic with a mass given by (\ref{massground}) and it is removed by the GSO projection.

The first set of physical states in the GSO projected spectrum are those created by a fermionic creation mode with index $r=\frac{1}{2}$. In particular those generated by the oscillators $\widetilde{\Psi^2}_{\!-\frac{1}{2}+\nu_a}$ and 
$\widetilde{\overbar{\Psi}^3}_{\!-\frac{1}{2}+\nu_a}$ are massless since the
energy carried by these modes exactly cancels that of the vacuum.
Therefore, the vertex operators at zero momentum associated to these right-moving
massless excitations in the $(-1)$-superghost picture are:
\begin{equation}
\begin{aligned}
\widetilde{\mathcal{V}}^1_a(\bar{z})&=-\widetilde{\sigma}_{a}(\bar{z})
:\!\widetilde{\Psi^2}(\bar{z})\,\widetilde{s}_{a}(\bar{z})\!:\rme^{-\widetilde{\phi}(\bar{z})}~,\\[1mm]
\widetilde{\mathcal{V}}^2_a(\bar{z})&=\phantom{-}\widetilde{\sigma}_{a}(\bar{z})
:\!\widetilde{\overbar{\Psi}^3}(\bar{z})\,\widetilde{s}_{a}(\bar{z})\!:\rme^{-
\widetilde{\phi}(\bar{z})}~.\phantom{\Big|}
\end{aligned}
\label{vertex2}
\end{equation}
These are conformal fields of weight 1 and we collectively denote them as
$\widetilde{\mathcal{V}}^\beta_a(\bar{z})$ with $\beta=1,2$. We point out that the $-$
sign in the first line above is introduced because in this way the two operators form
a doublet transforming in the spinor representation of SU(2)$_+$, as explained in
Appendix~\ref{SO(4)}.

\subsubsection{Sectors with twist parameter $(1-\nu_a) >\frac{1}{2}$}

Apart from a few subtleties, the conclusions obtained in the previous subsection for the twisted
sectors with $\nu_a<\frac{1}{2}$, are 
valid also in the twisted sectors with $(1-\nu_a) >\frac{1}{2}$ provided one exchanges the role of the complex directions 2 and 3, and uses the sector label $(M-a)$. Thus, we can rather brief in our presentation.

In the left-moving part, 
the fermionic creation modes are the oscillators ${\Psi}^{2}_{-r+\nu_a}$, $\overbar{\Psi}^{2}_{-r-\nu_a}$, 
${\Psi}^{3}_{-r-\nu_a}$ and $\overbar{\Psi}^{3}_{-r+\nu_a}$ where $r$ is a positive half-integer.
They act on the twisted vacuum $|\Omega_{M-a}\rangle$ which is defined by
\begin{equation}
|\Omega_{M-a}\rangle =\lim_{z\to 0}\,\sigma_{M-a}(z)\,s_{M-a}(z)\,|0\rangle
\end{equation}
with
\begin{equation}
\begin{aligned}
\sigma_{M-a}(z)&=\sigma_{1-\nu_a}^2(z)\,\sigma_{\nu_a}^3(z)~,\quad\text{and}\quad 
s_{M-a}(z)=s_{-\nu_a}^2(z)\,s_{\nu_a}^3(z)~.
\end{aligned}
\label{sigmas3}
\end{equation}
The ground state $|\Omega_{M-a}\rangle$ is tachyonic with a mass given by 
(\ref{massground}) and is removed by the GSO projection.
At the first excited level, instead, we find two massless states created by the oscillators
${\Psi}^{2}_{-\frac{1}{2}+\nu_a}$ and $\overbar{\Psi}^{3}_{-\frac{1}{2}+\nu_a}$, which 
correspond to the following vertex operators at zero momentum:
\begin{equation}
\begin{aligned}
\mathcal{V}^1_{M-a}(z)&=-\sigma_{M-a}(z):\!{\Psi}^{2}(z)\,s_{M-a}(z)\!:\rme^{-\phi(z)}~,\\[1mm]
\mathcal{V}^2_{M-a}(z)&=\phantom{-}\sigma_{M-a}(z):\!\overbar{\Psi}^{3}(z)\,s_{M-a}(z)\!:\rme^{-\phi(z)}~.
\end{aligned}
\label{vertex3}
\end{equation}
These are conformal fields of weight 1 which we collectively denote as 
$\mathcal{V}^\alpha_{M-a}(z)$ with $\alpha=1,2$. Again the $-$ sign in the first line
is inserted so that these two operators transform as a doublet in the spinor representation
of SU(2)$_+$ (see Appendix~\ref{SO(4)}).

Finally, in the right-moving part
the oscillators $\widetilde{\Psi^2}_{\!-r-\nu_a}$,
$\widetilde{\overbar{\Psi}^2}_{\!-r+\nu_a}$, $\widetilde{\Psi^3}_{\!-r+\nu_a}$ and $\widetilde{\overbar{\Psi}^3}_{\!-r-\nu_a}$ where $r$ is a positive half-integer, are creation modes. They
act on the twisted vacuum defined by
\begin{equation}
|\widetilde{\Omega}_{M-a}\rangle =\lim_{\bar{z}\to 0}\,\widetilde{\sigma}_{M-a}(\bar{z})
\,\widetilde{s}_{M-a}(\bar{z})\,|0\rangle
\end{equation}
where 
\begin{equation}
\begin{aligned}
\widetilde{\sigma}_{M-a}(\bar{z})& =\,\widetilde{\sigma}_{\nu_a}^{\,2}(\bar{z})\,
\widetilde{\sigma}_{1-\nu_a}^{\,3}(\bar{z})~,\quad\text{and}\quad 
\widetilde{s}_{M-a}(\bar{z})=\,\widetilde{s}_{\nu_a}^{\,2}(\bar{z})\,\widetilde{s}_{-\nu_a}^{\,3}(\bar{z})~.
\end{aligned}
\label{sigmas4}
\end{equation}
As before this vacuum state is tachyonic and removed by the GSO projection. 
On the other hand, the states created by $\widetilde{\overbar{\Psi}^2}_{\!-\frac{1}{2}+\nu_a}$ and  $\widetilde{\Psi^3}_{\!-\frac{1}{2}+\nu_a}$ are massless and selected by the GSO projection. They correspond to the following
vertex operators at zero momentum:
\begin{equation}
\begin{aligned}
\widetilde{\mathcal{V}}^1_{M-a}(\bar{z})&=\widetilde{\sigma}_{M-a}(\bar{z})
:\!\widetilde{\Psi}^{3}(\bar{z})\,\widetilde{s}_{M-a}(\bar{z})\!:\rme^{-\widetilde{\phi}(\bar{z})}~,\\[1mm]
\widetilde{\mathcal{V}}^2_{M-a}(\bar{z})&=\widetilde{\sigma}_{M-a}(\bar{z})
:\!\widetilde{\overbar{\Psi}^2}(\bar{z})\,\widetilde{s}_{M-a}(\bar{z})\!:\rme^{-
\widetilde{\phi}(\bar{z})}~,
\end{aligned}
\label{vertex4}
\end{equation}
which are conformal fields of weight 1. We collectively denote these vertex operators as
$\widetilde{\mathcal{V}}^\beta_{M-a}(\bar{z})$ with $\beta=1,2$, since they transform
as a doublet of SU(2)$_+$ (see Appendix~\ref{SO(4)}).

We summarize our results on the massless vertex operators of the twisted NS sectors in Table \ref{tab:table5} below. 

\begin{table}[ht]
  \begin{center}
    \begin{tabular}{|c|c|} \hline
     \text{~Vertex operator} \phantom{\Big|}&\text{State} \\
      \hline\hline
 $\phantom{\bigg|} \mathcal{V}^1_a(z)=\sigma_{a}(z)\, 
 :\!{\Psi}^{3}(z)\,s_{a}(z) \!:\rme^{-\phi(z)}$ & $\phantom{\bigg|}$ ${\Psi}^{3}_{-\frac{1}{2}+\nu_a}\,|\Omega_{a}\rangle_{(-1)}$ \\[3mm]
 $\phantom{\bigg|} \mathcal{V}^2_a(z)=
 \sigma_{a}(z)\,
 :\!\overbar{\Psi}^{2}(z)\,s_{a}(z) \!:\rme^{-\phi(z)}$& $\phantom{\bigg|}$ $\overline{\Psi}^{2}_{-\frac{1}{2}+\nu_a}\,|\Omega_{a}\rangle_{(-1)}$
 \\[3mm]\hline
$\phantom{\bigg|} \widetilde{\mathcal{V}}^1_a(\bar{z})=-
\widetilde{\sigma}_{a}(\bar{z})\, 
:\!\widetilde{\Psi^2}(\bar{z})\,\widetilde{s}_{a}(\bar{z})\!:\rme^{-\widetilde{\phi}(\bar{z})}$ &
$-\widetilde{\Psi^2}_{-\frac{1}{2}+\nu_a}\,|\widetilde{\Omega}_{a}\rangle_{(-1)}$
\\[3mm]
 $\phantom{\bigg|} \widetilde{\mathcal{V}}^2_a(\bar{z})=
 \widetilde{\sigma}_{a}(\bar{z})\, 
:\!\widetilde{\overbar{\Psi}^3}(\bar{z})\,\widetilde{s}_{a}(\bar{z})\,\!:\rme^{-\widetilde{\phi}(\bar{z})}$ & $\widetilde{\overbar{\Psi}^3}_{-\frac{1}{2}+\nu_a}\,|\widetilde{\Omega}_{a}\rangle_{(-1)}$\\[3mm]
\hline\hline
$\phantom{\bigg|} \mathcal{V}^1_{M-a}(z)=-\sigma_{M-a}(z)\, 
 :\!{\Psi}^{2}(z)\,s_{M-a}(z)\!:\rme^{-\phi(z)}$ & $\phantom{\bigg|}$ $-{\Psi}^{2}_{-\frac{1}{2}+\nu_a}\,|\Omega_{M-a}\rangle_{(-1)}$ \\[3mm]
 $\phantom{\bigg|} \mathcal{V}^2_{M-a}(z)=
 \sigma_{M-a}(z)\, 
 :\!\overbar{\Psi}^{\,3}(z)\,s_{M-a}(z)\!:\rme^{-\phi(z)}$& $\phantom{\bigg|}$ $\overbar{\Psi}^{3}_{-\frac{1}{2}+\nu_a}\,|\Omega_{M-a}\rangle_{(-1)}$
 \\[3mm]\hline
 $\phantom{\bigg|} \widetilde{\mathcal{V}}^1_{M-a}(\bar{z})=
\widetilde{\sigma}_{M-a}(\bar{z})\, 
:\!\widetilde{\Psi^3}(\bar{z})\,\widetilde{s}_{M-a}(\bar{z})\!:\rme^{-\widetilde{\phi}(\bar{z})}$ &
$\widetilde{\Psi^3}_{-\frac{1}{2}+\nu_a}\,|\widetilde{\Omega}_{M-a}\rangle_{(-1)}$
\\[3mm]
 $\phantom{\bigg|} \widetilde{\mathcal{V}}^2_{M-a}(\bar{z})=
 \widetilde{\sigma}_{M-a}(\bar{z})\, 
:\!\widetilde{\overbar{\Psi}^2}(\bar{z})\,\widetilde{s}_{M-a}(\bar{z})\!:\rme^{-\widetilde{\phi}(\bar{z})}$ & $\widetilde{\overbar{\Psi}^2}_{-\frac{1}{2}+\nu_a}\,|\widetilde{\Omega}_{M-a}\rangle_{(-1)}$\\
\hline
    \end{tabular}
  \end{center}
  \caption{The vertex operators and the corresponding states in the left- and right-moving parts of
  the various twisted NS sectors. Here the label $a$ takes values in the range $\big[1, \frac{M-1}{2}\big]$, and in the last column the subscript $(-1)$ on the kets identifies the superghost picture.}
    \label{tab:table5}
\end{table}

\subsubsection{Two-point functions in the twisted NS sectors}
\label{NStwistcorrelators} 
Given the explicit form of the vertex operators that we have derived, it is rather straightforward to
compute their two-point functions. As a first step, we observe that there are no 
non-vanishing correlators between left (or right) operators of the same twisted sector, due to 
the presence of the bosonic twist fields; in fact for any complex direction $j$ one has
\cite{Dixon:1986qv}
\begin{equation}
\big\langle\sigma_{\nu_a}^j(z_1)\,\sigma_{\nu_b}^j(z_2)\big\rangle 
=\frac{\delta_{\nu_b,1-\nu_a}}{
(z_1-z_2)^{\nu_a(1-\nu_a)}}~,
\label{sigmasigma}
\end{equation} 
and similarly in the right sector. This implies that only the correlator $\langle \sigma_a(w_1)\,\sigma_{M-a}(w_2)\rangle$ is non vanishing. Therefore, only the two-point functions 
between vertex operators in sectors $a$ and $(M-a)$ are non-zero. Another important point
to consider is that these vertex operators inherit the fermionic statistics from the fermionic fields
that are present in their definitions.

Let us then compute the two-point function between $\mathcal{V}_a^1$ and $\mathcal{V}_{M-a}^2$. Using
(\ref{sigmasigma}) and the basic conformal field theory correlators
\begin{equation}
\begin{aligned}
\big\langle \!:\!\Psi^3(z_1)\,s_a(z_1)\!:\,\,:\!\overbar{\Psi}^3(z_2)\,s_{M-a}(z_2)\!: \!\big\rangle &=
\frac{1}{(z_1-z_2)^{1-2\nu_a(1-\nu_a)}}~,\\[1mm]
\big\langle\rme^{-\phi(z_1)}\,\rme^{-\phi(z_2)}\big\rangle &=\frac{1}{z_1-z_2}~,
\end{aligned}
\end{equation}
we obtain
\begin{equation}
\big\langle \mathcal{V}^{1}_a(z_1)\,\mathcal{V}^{2}_{M-a}(z_2)\big\rangle 
=\frac{1}{(z_1-z_2)^2}~.
\label{v1v2}
\end{equation}
In a similar way, using
\begin{equation}
\big\langle \!:\!\overbar{\Psi}^2(z_1)\,s_a(z_1)\!:\,\,:\!{\Psi}^2(z_2)\,s_{M-a}(z_2)\!: \!\big\rangle =
\frac{1}{(z_1-z_2)^{1-2\nu_a(1-\nu_a)}}~,
\end{equation}
and taking into account the explicit negative sign in $\mathcal{V}^1_{M-a}$, we get
\begin{equation}
\big\langle \mathcal{V}^{2}_a(z_1)\,\mathcal{V}^{1}_{M-a}(z_2)\big\rangle 
=\frac{-1}{(z_1-z_2)^2}~.
\label{v2v1}
\end{equation}
Furthermore, the two-point functions between $\mathcal{V}^1_a$ and $\mathcal{V}^1_{M-a}$ and between
$\mathcal{V}^2_a$ and $\mathcal{V}^2_{M-a}$ vanish since their fermionic charges do not match.
Thus, altogether, we have
\begin{equation}
\big\langle \mathcal{V}^{\alpha}_a(z_1)\,\mathcal{V}^{\beta}_{M-a}(z_2)\big\rangle 
=\frac{(\epsilon^{-1})^{\alpha\beta}}{(z_1-z_2)^2}
\label{vv}
\end{equation}
where we have defined 
\begin{equation}
\epsilon=\begin{pmatrix}
0&-1\\
1&0\end{pmatrix}~.
\label{epsilon}
\end{equation}
By taking into account the fermionic statistics of
the vertex operators and the anti-symmetry of $\epsilon$, we also find
\begin{equation}
\big\langle \mathcal{V}^{\alpha}_{M-a}(z_1)\,\mathcal{V}^{\beta}_{a}(z_2)\big\rangle 
=\frac{(\epsilon^{-1})^{\alpha\beta}}{(z_1-z_2)^2}~.
\label{vvM}
\end{equation}
Notice that (\ref{vv}) and (\ref{vvM}) may be unified in a single formula by promoting the index $a$ to
the complete index $\widehat{a}$. This shows that despite the differences in the structure of the
states and vertex operators in the fermionic formalism, all twisted sectors are actually treated on equal  footing.

Similarly, in the right-moving sector, we obtain
\begin{equation}
\big\langle \widetilde{\mathcal{V}}^{\alpha}_{M-a}(\bar{z}_1)
\,\widetilde{\mathcal{V}}^{\beta}_{a}(\bar{z}_2)\big\rangle 
=\big\langle \widetilde{\mathcal{V}}^{\alpha}_{a}(\bar{z}_1)
\,\widetilde{\mathcal{V}}^{\beta}_{M-a}(\bar{z}_2)\big\rangle 
=\frac{(\epsilon^{-1})^{\alpha\beta}}{(\bar{z}_1-\bar{z}_2)^2}~.
\end{equation}
{From} these two-point functions it is possible to infer the conjugate vertex operators as follows:
\begin{equation}
\begin{aligned}
\big(\mathcal{V}_{M-a}(z)\big)^\dagger_\alpha=\mathcal{V}_{a}^\beta(z)\,\epsilon_{\beta\alpha}~&,~~
\big(\mathcal{V}_a(z)\big)^\dagger_\alpha=\mathcal{V}_{M-a}^\beta(z)\,\epsilon_{\beta\alpha}~,\\[1mm]
\big(\widetilde{\mathcal{V}}_{a}(z)\big)^\dagger_\alpha=\widetilde{V}_{M-a}^\beta(z)\,\epsilon_{\beta\alpha}~&,~~
\big(\widetilde{\mathcal{V}}_{M-a}(z)\big)^\dagger_\alpha=\widetilde{V}_{a}^\beta(z)\,\epsilon_{\beta\alpha}~.
\end{aligned}
\label{daggerNS}
\end{equation}

\subsection{The massless NS/NS vertex operators}
\label{NSNStwisted}

The massless closed string excitations in the twisted NS/NS sectors are obtained by combining the left- and right-moving massless states that we have obtained in the previous subsection.
In the sectors with twist parameter $\nu_a < \frac{1}{2}$, 
they are then described by the following vertex operators at zero momentum
\begin{align}
b^{(a)}_{\alpha\beta}\,\,\mathcal{V}_a^\alpha(z)\,\widetilde{
\mathcal{V}}^\beta_a(\bar{z})
\label{vertexopsVba}
\end{align}
where $b^{(a)}_{\alpha\beta}$ are four constant complex fields. 

Similarly, in the sectors with twist parameter $(1-\nu_a)>\frac{1}{2}$, the massless closed string excitations are described by the vertex operators at zero momentum
\begin{align}
b^{(M-a)}_{\alpha\beta}\,\,\mathcal{V}_{M-a}^\alpha(z)\,\widetilde{\mathcal{V}}^\beta_{M-a}(\bar{z})
\label{vertexopsVbM-a}
\end{align}
where again $b^{(M-a)}_{\alpha\beta}$ are four constant complex fields. 

The constants $b^{(a)}$ and $b^{(M-a)}$ can be considered as a background in which the 
string theory on the orbifold is defined. Given the structure of the vertex operators
there are non-trivial relations among them.
In particular, using (\ref{daggerNS}) one finds that
\begin{equation}
\Big(b^{(a)}_{\alpha\beta}\,
\mathcal{V}^{\alpha}_a(z) \,\widetilde{\mathcal{V}}^{\beta}_a(\bar{z})\Big)^\dagger
=b^{(M-a)}_{\alpha\beta}\,\mathcal{V}^{\alpha}_{M-a}(z)\,
\widetilde{\mathcal{V}}^{\beta}_{M-a}(\bar{z})
\end{equation}
where
\begin{equation}
b^{(M-a)}_{11}=-b^{(a)\,\star}_{22}~,\quad
b^{(M-a)}_{12}=b^{(a)\,\star}_{21}~,\quad
b^{(M-a)}_{21}=b^{(a)\,\star}_{12}~,\quad
b^{(M-a)}_{22}=-b^{(a)\,\star}_{11}~,
\label{starNSexp}
\end{equation}
or, equivalently in matrix notation,
\begin{equation}
b^{(M-a)}=\epsilon\,b^{(a)\,\star}\,\epsilon~.
\label{starNS}
\end{equation}
These relations, which also appear in \cite{Douglas:1996sw}, show that if one turns on background values for the closed string fields in the twisted sector $a$, one also turns on background values for the fields in the twisted sector $(M-a)$ and $\emph{viceversa}$,
in such a way that the total background configuration is real.

\subsection{Twisted R sectors}
\label{twistR}
The $\mathbb{Z}_M$ orbifold (\ref{orbifold}) breaks the isometry of the ten-dimensional 
space as follows: 
\begin{equation}
\mathrm{SO}(10)~\longrightarrow~ \mathrm{SO}(6)\times\mathrm{SO}(2)\times \mathrm{SO}(2)~,
\end{equation}
where SO$(6)$ acts on the first, fourth and fifth complex directions, which are not affected 
by the orbifold action. Correspondingly, the untwisted vacuum of the R sector which carries the 
32-dimensional spinor representation of SO(10) decomposes into eight massless spinors of 
SO(6). Four of these are chiral and four anti-chiral. We denote the four chiral vacuum states by
\begin{equation}
\Big|A,\pm\frac{1}{2},\pm\frac{1}{2}\Big\rangle
\label{A}
\end{equation}
where $A\in\mathbf{4}$ labels the four different components of the chiral spinor representation
of SO(6) and the four pairs of $\pm\frac{1}{2}$ denote the spinor weights along the second
and third complex directions where the orbifold acts. Similarly, the four anti-chiral vacuum states
are denoted by 
\begin{equation}
\Big|\dot{A},\pm\frac{1}{2},\pm\frac{1}{2}\Big\rangle
\label{Adot}
\end{equation}
where $\dot{A}\in\mathbf{\bar{4}}$ spans the four-dimensional anti-chiral spinor representation of SO(6).

In the twisted R sectors, not all such chiral and anti-chiral states remain massless. Indeed, the
fermionic twist fields change the spinor weights in the orbifolded directions, so that conformal dimensions and the GSO parities of the corresponding vertex operators are modified. In the following we present a brief description of the spectrum in the
various twisted R sectors, focusing on the massless excitations.

\subsubsection{Sectors with twist parameter $\nu_a<\frac{1}{2}$}

In these sectors the left-moving bosonic and fermionic twist fields $\sigma_a$ and $s_a$
are given in (\ref{sigmas}).
When we act with $s_{a}$ on the states (\ref{A}) and (\ref{Adot}), the charges
in the directions 2 and 3 become
\begin{equation}
\varepsilon_2= \pm \frac{1}{2} + \nu_a\quad\mbox{and} 
\quad \varepsilon_3=\pm\frac{1}{2} - \nu_a
\label{efft}
\end{equation}
depending on their initial values. Because of this, not all choices of signs lead to massless
configurations. In fact, the mass vanishes only if 
\begin{equation}
\varepsilon_2^2= \varepsilon_3^2 = \Big(\frac{1}{2} - \nu_a\Big)^2~.
\label{efftt}
\end{equation}
Combining this with (\ref{efft}), we see that the only solution is
\begin{equation}
\varepsilon_2= -\varepsilon_3 = -\frac{1}{2} + \nu_a~,
\end{equation}
so that, instead of $s_a(z)$, we can consider the effective fermionic twist
\begin{equation}
r_a(z) = s^2_{\nu_a-\frac{1}{2}}(z)\,s^3_{-\nu_a+\frac{1}{2}}(z)
\label{raeff}
\end{equation}
which is a conformal field of weight $\big(\frac{1}{4}-\nu_a(1-\nu_a)\big)$.

In the R sector, there are two fundamental superghost pictures that one considers: the 
$(-\frac{1}{2})$- and the $(-\frac{3}{2})$-pictures \cite{Friedan:1985ge}. Enforcing the GSO
projection, in the $(-\frac{1}{2})$-picture one selects the chiral spinor of SO(6), while in
the $(-\frac{3}{2})$-picture one selects the anti-chiral one. In this way, in fact, the sum of
the spinor weights minus the superghost-charge is always an even integer. Thus
we are led to introduce the following two vertex operators at zero momentum
\begin{subequations}
\begin{align}
\mathcal{V}^A_a(z)&=\sigma_{a}(z)\,r_a(z)\, S^A(z) \rme^{-\frac{1}{2}\phi(z)}~,
\label{vertexR1}\\
\mathcal{V}^{\dot{A}}_a(z)&=\sigma_{a}(z) \,r_a (z)\,S^{\dot{A}}(z) \rme^{-\frac{3}{2}\phi(z)}~,
\label{vertexR11}
\end{align}%
\end{subequations}
where $S^A$ and $S^{\dot{A}}$ are, respectively, the chiral and anti-chiral spin-fields of
SO(6) \cite{Friedan:1985ge,Kostelecky:1986xg}. Both vertex operators are conformal fields of weight 1 and define the following massless twisted vacuum states:
\begin{equation}
\begin{aligned}
|A_a\rangle_{(-\frac{1}{2})} &= \lim_{z\to 0} \,\mathcal{V}^{A}_a(z)\,|0\rangle~,\\
|\dot{A}_a\rangle_{(-\frac{3}{2})} &= \lim_{z\to 0} \,\mathcal{V}^{\dot{A}}_a(z)\,|0\rangle~.
\end{aligned}
\label{Rstates1}
\end{equation}

As far as the right-moving part is concerned, the bosonic and fermionic twist fields are given
in (\ref{sigmasr}). Therefore, we can repeat the previous analysis by simply replacing everywhere
$\nu_a$ with $(1-\nu_a)$. In this way we find the following two physical vertex operators
of weight 1:
\begin{subequations}
\begin{align}
\widetilde{\mathcal{V}}^{A}_a(\bar{z})&=\widetilde{\sigma}_{a}(\bar{z})
\, \widetilde{r}_{a}(\bar{z})\,\widetilde{S}^{A}(\bar{z})\,\rme^{-\frac{1}{2}\widetilde{\phi}(\bar{z})}
~,\\
\widetilde{\mathcal{V}}^{\dot{A}}_a(\bar{z})&=
\widetilde{\sigma}_{a}(\bar{z})\,\widetilde{r}_{a}(\bar{z})\,
\widetilde{S}^{\dot{A}}(\bar{z})\,\rme^{-\frac{3}{2}\widetilde{\phi}(\bar{z})}~,
\end{align}%
\label{vertexR2}
\end{subequations}
where the effective fermionic twist is given by
\begin{equation}
\widetilde r_a(\bar{z}) = \widetilde s^2_{-\nu_a+\frac{1}{2}}(\bar{z})
\,\widetilde s^3_{\nu_a-\frac{1}{2}}(\bar{z})~.
\label{raefft}
\end{equation}
The massless states corresponding to these vertex operators are
\begin{equation}
\begin{aligned}
|\widetilde{A}_a\rangle_{(-\frac{1}{2})} &= \lim_{\bar{z}\to 0} \,
\widetilde{\mathcal{V}}^{A}_a(\bar{z})\,|0\rangle~,\\
|\widetilde{\dot{A}}_a\rangle_{(-\frac{3}{2})} &= \lim_{\bar{z}\to 0} \,
\widetilde{\mathcal{V}}^{\dot{A}}_a(\bar{z})\,|0\rangle~.
\end{aligned}
\label{Rstates2}
\end{equation}

\subsubsection{Sectors with twist parameter $(1-\nu_a) > \frac{1}{2}$}
These sectors can be described in the same manner as before by simply exchanging the roles
of the complex directions 2 and 3, and using $(M-a)$ as twist label. 
Thus, we merely present the physical GSO projected massless vertex operators at zero momentum. 
In the left-moving part they are
\begin{subequations}
\begin{align}
\mathcal{V}^{A}_{M-a}(z)&=\sigma_{M-a}(z)\,r_{M-a}(z)\,S^{A}(z)\,\rme^{-\frac{1}{2}\phi(z)}~,\label{vertexR31}\\
\mathcal{V}^{\dot{A}}_{M-a}(z)&=\sigma_{M-a}(z)\,r_{M-a}(z)\,S^{\dot{A}}(z)\,
\rme^{-\frac{3}{2}\phi(z)}~,
\label{vertexR311}
\end{align}%
\end{subequations}
with
\begin{equation}
r_{M-a}(z) =s^2_{-\nu_a+\frac{1}{2}}(z)\,s^3_{\nu_a-\frac{1}{2}}(z)~.
\label{raeffm}
\end{equation}
In the right-moving part, instead, they are
\begin{subequations}
\begin{align}
\widetilde{\mathcal{V}}^{A}_{M-a}(\bar{z})&=\widetilde{\sigma}_{M-a}(\bar{z})
\,\widetilde{r}_{M-a}(\bar{z})\,\widetilde{S}^{A}(\bar{z})\,
\rme^{-\frac{1}{2}\widetilde{\phi}(\bar{z})}~,\\
\widetilde{\mathcal{V}}^{\dot{A}}_{M-a}(\bar{z})&
=\widetilde{\sigma}_{M-a}(\bar{z})\,\widetilde{r}_{M-a}(\bar{z})\,
\widetilde{S}^{\dot{A}}(\bar{z})\,\rme^{-\frac{3}{2}\widetilde{\phi}(\bar{z})}~.
\end{align}
\label{vertexR4}
\end{subequations}
where
\begin{equation}
\widetilde r_{M-a}(\bar{z}) = \widetilde s^2_{\nu_a-\frac{1}{2}}(\bar{z})\,
\widetilde s^3_{-\nu_a+\frac{1}{2}}(\bar{z})~.
\label{raeffm-a}
\end{equation}
When acting on the Fock vacuum these vertex operators create the twisted ground states which have
the same expressions as in (\ref{Rstates1}) and (\ref{Rstates2}) with the obvious changes in notation.

We summarize our findings in Table~\ref{tab:table6} below. 

\begin{table}[ht]
  \begin{center}
    \begin{tabular}{|c|c|} \hline
     \text{~Vertex operator} \phantom{\Big|}&\text{State} \\
      \hline\hline
 $\phantom{\bigg|} \mathcal{V}^A_a(z)=\sigma_{a}(z)\,r_a(z)\, S^A(z) \rme^{-\frac{1}{2}\phi(z)}$ & $\phantom{\bigg|}$ $|A_a\rangle_{(-\frac{1}{2})}$ \\[3mm]
 $\phantom{\bigg|} \mathcal{V}^{\dot{A}}_a(z)=\sigma_{a}(z) \,r_a (z)\,S^{\dot{A}}(z) 
 \rme^{-\frac{3}{2}\phi(z)}$& 
 $\phantom{\bigg|}$ $|\dot{A_a}\rangle_{(-\frac{3}{2})}$
 \\[3mm]\hline
$\phantom{\bigg|} \widetilde{\mathcal{V}}^{A}_a(\bar{z})=\widetilde{\sigma}_{a}(\bar{z})
\, \widetilde{r}_{a}(\bar{z})\,\widetilde{S}^{A}(\bar{z})\,\rme^{-\frac{1}{2}\widetilde{\phi}(\bar{z})}$ 
&$|\widetilde{A}_a\rangle_{(-\frac{1}{2})} $
\\[3mm]
 $\phantom{\bigg|} \widetilde{\mathcal{V}}^{\dot{A}}_a(\bar{z})
=\widetilde{\sigma}_{a}(\bar{z})\,\widetilde{r}_{a}(\bar{z})\,
\widetilde{S}^{\dot{A}}(\bar{z})\,\rme^{-\frac{3}{2}\widetilde{\phi}(\bar{z})}$ & 
$|\widetilde{\dot{A}}_a\rangle_{(-\frac{3}{2})}$\\[3mm]
\hline\hline
$~~~\phantom{\bigg|} \mathcal{V}^{A}_{M-a}(z)=\sigma_{M-a}(z)\,r_{M-a}(z)\,S^{A}(z)
\,\rme^{-\frac{1}{2}\phi(z)}~~~$ & 
$~~~\phantom{\bigg|}$ $|A_{M-a}\rangle_{(-\frac{1}{2})}~~$ \\[3mm]
 $~~~\phantom{\bigg|}\mathcal{V}^{\dot{A}}_{M-a}(z)=\sigma_{M-a}(z)\,r_{M-a}(z)
 \,S^{\dot{A}}(z)\,\rme^{-\frac{3}{2}\phi(z)}~~~$& 
 $~~~\phantom{\bigg|}$ $|\dot{A}_{M-a}\rangle_{(-\frac{3}{2})} ~~$
 \\[3mm]\hline
 $~~~\phantom{\bigg|} \widetilde{\mathcal{V}}^{A}_{M-a}(\bar{z})=\widetilde{\sigma}_{M-a}(\bar{z})
\,\widetilde{r}_{M-a}(\bar{z})\,\widetilde{S}^{A}(\bar{z})\,\rme^{-\frac{1}{2}\widetilde{\phi}(\bar{z})}~~~$ &
$~~~|\widetilde{A}_{M-a}\rangle_{(-\frac{1}{2})}~~$ \\[3mm]
$~~~\phantom{\bigg|} \widetilde{\mathcal{V}}^{\dot{A}}_{M-a}(\bar{z})=\widetilde{\sigma}_{M-a}(\bar{z})
\,\widetilde{r}_{M-a}(\bar{z})\,\widetilde{S}^{\dot{A}}(\bar{z})\,\rme^{-\frac{3}{2}\widetilde{\phi}(\bar{z})}~~~$ &
$~~~|\widetilde{\dot{A}}_{M-a}\rangle_{(-\frac{3}{2})}~~$ \\
\hline
    \end{tabular}
  \end{center}
  \caption{The vertex operators and the corresponding states in the left- and right-moving parts of
  the twisted R sectors.}
    \label{tab:table6}
\end{table}

\subsubsection{Two-point functions in the twisted R sectors}

As we have seen in the twisted NS sectors, the only non-vanishing two-point functions 
necessarily involve the left-moving (or right-moving) vertex operators in 
complementary sectors $a$ and $(M-a)$, because of the two-point functions (\ref{sigmasigma}).
Of course, the same is true in the twisted R sectors. Furthermore, in order to soak up the background charge in the superghost sector, only the overlaps between states in the $(-\frac{1}{2})$- 
and $(-\frac{3}{2})$-pictures, or {\emph{viceversa}}, are non-zero. Taking this into account
and using standard results from conformal field theory, we find
\begin{equation}
\begin{aligned}
\big\langle \mathcal{V}^{A}_a(z_1)\,\mathcal{V}^{\dot B}_{M-a}(z_2)\big\rangle &
=\big\langle \mathcal{V}^{A}_{M-a}(z_1)\,\mathcal{V}^{\dot B}_a(z_2)\big\rangle
=\frac{(C^{-1})^{A \dot B}}{(z_1-z_2)^2}~,\\[1mm]
\big\langle \mathcal{V}^{\dot A}_a(z_1)\,\mathcal{V}^{B}_{M-a}(z_2)\big\rangle &
=\big\langle \mathcal{V}^{\dot A}_{M-a}(z_1)\,\mathcal{V}^{B}_a(z_2)\big\rangle =
\frac{(C^{-1})^{\dot A  B}}{(z_1-z_2)^2}~,
\end{aligned}
\label{2pointR}
\end{equation}
where $C$ is the charge conjugation matrix of SO(6) (see Appendix~\ref{twistedRspinors}).
Of course, analogous correlators hold for the right-moving vertex operators. 

{From} the first line of (\ref{2pointR}), we read the following conjugation rules 
\begin{equation}
\begin{aligned}
\big(\mathcal{V}_{M-a}(z)\big)^\dagger_{\dot{B}}&=\mathcal{V}^{A}_{a}(z)\, C_{A\dot{B}}~,\\
\big(\mathcal{V}_{a}(z)\big)^\dagger_{\dot{B}}&=\mathcal{V}^{A}_{M-a}(z)\, C_{A\dot{B}}~,
\end{aligned}
\label{leftrelR1}
\end{equation}
while from the second line we obtain the same relations with dotted and undotted indices exchanged. The same formulas apply also for the right-moving vertices.

\subsection{The massless R/R vertex operators}
The massless closed string excitations in the twisted R/R sectors are obtained by combining
left and right movers. We shall work with the asymmetric superghost pictures
$(-\frac{1}{2},-\frac{3}{2})$ or $(-\frac{3}{2},-\frac{1}{2})$, so that the corresponding closed string fields are R/R potentials. In the twisted sector labeled by $a$ we choose the 
$(-\frac{1}{2},-\frac{3}{2})$-picture and write the following massless 
vertex operators at zero momentum:\,%
\footnote{In \cite{Billo:1998vr} it is shown that the complete BRST invariant vertex operators in the asymmetric superghost pictures are an infinite sum of terms characterized by the number of
superghost zero modes. For our purposes, however, only the first (and simplest) 
terms in these sums is relevant since all the others decouple and thus can be discarded.}. 
\begin{align}
\mathcal{C}^{(a)}_{A\dot{B}}\,\mathcal{V}_a^A(z)\,\,\widetilde{\mathcal{V}}^{\dot{B}}_a(\bar{z})
\label{RR13}
\end{align}
where $\mathcal{C}^{(a)}_{A\dot{B}}$ are sixteen constant complex fields. These constants can 
be considered as a background in which the orbifold closed string theory is defined. 

In the twisted sector labeled by $(M-a)$ we choose, instead, the other asymmetric superghost
picture, namely the $(-\frac{3}{2},-\frac{1}{2})$-picture, and consider the following massless
vertex operators
\begin{align}
\mathcal{C}^{(M-a)}_{\dot{A}{B}}\,
\mathcal{V}_{M-a}^{\dot{A}}(z)\,\,\widetilde{\mathcal{V}}^{B}_{M-a}(\bar{z})
\label{RR14}
\end{align}
where $\mathcal{C}^{(M-a)}_{A\dot{B}}$ are other sixteen constant complex fields contributing to
the background in which the closed string propagates.

Notice that in writing the vertex operators (\ref{RR13}) and (\ref{RR14}) for the twisted 
R/R potentials, we have correlated the choice of picture numbers with the twisted sector. 
Of course we could have made different choices, but they would lead to the same results. 
In fact, it is well-known that in a BRST invariant framework like ours, the way in which the superghost pictures are distributed is completely arbitrary, provided one satisfies the global 
constraints due to the presence of a background charge, and that the physical results do not 
depend on this choice. However, our picture assignment is particularly convenient because it immediately implies that the R/R potentials in the $a$-th twisted sector are naturally 
related to those in the sector $(M-a)$ by complex conjugation, exactly as it happens in the 
twisted NS/NS sectors. Indeed, we have
\begin{equation}
\begin{aligned}
\Big(\mathcal{C}^{(a)}_{A\dot{B}}\,\mathcal{V}_a^{A}(z)\,
\widetilde{\mathcal{V}}^{\dot{B}}_{a}(\bar{z})\Big)^{\dagger}
=
\mathcal{C}^{(M-a)}_{\dot{A}{B}}\,\mathcal{V}_{M-a}^{\dot{A}}(z)\,
\widetilde{\mathcal{V}}^{B}_{M-a}(\bar{z})~,
\end{aligned}
\end{equation}
where, in matrix notation,
\begin{equation}
\mathcal{C}^{(M-a)}=C\,\mathcal{C}^{(a)\,\star}\,C~,
\label{starR}
\end{equation}
which is the strict analogue of (\ref{starNS}) holding in the NS/NS sectors. We therefore see that 
by turning on a R/R background potential value in the twisted sector $a$, one also turns on a background R/R potential in the twisted sector $(M-a)$ and $\emph{viceversa}$,
in such a way that the total configuration is real.

\section{Fractional D3-branes in the ${\mathbb Z}_M$ orbifold}
\label{Dbranes}

We now turn to discuss the open strings in the 
${\mathbb Z}_M$ orbifold with the aim of analyzing surface defects in 4$d$ gauge theories
engineered on stacks of (fractional) D3-branes. As is well-known, a 
D-brane introduces a boundary on the 
string world-sheet where non-trivial relations between the left and the right movers of the closed strings take place. We will investigate these relations using the boundary state 
formalism (for a review, see for example \cite{DiVecchia:1999mal,DiVecchia:1999fje}) 
and then will analyze the massless open string spectrum on the brane world-volume.
Since our ultimate goal is to recover a string theory description of the surface defects in a 4$d$
gauge theory, we place the (fractional) D3-branes in such a way that they are partially extended
along the orbifold as originally proposed in \cite{Kanno:2011fw}. 
More precisely, we take the D3-brane world-volume to be 
$\mathbb{C}_{(1)}\times\mathbb{C}_{(2)}$ in such a way that the orbifold action breaks the 4$d$
Poincar\'e symmetry leaving unbroken the one in the first complex direction along which the surface defect is extended.

\subsection{Boundary states and reflection rules}
\label{sec:reflectionrules}

In the $\mathbb{Z}_M$ orbifold there are $M$ different types of fractional D-branes, 
labeled by an index $I=0,1,\ldots,M-1$, corresponding to the $M$ irreducible representations
of $\mathbb{Z}_M$. A fractional D3-brane of type $I$ can be described by a boundary state which contains an
untwisted component $|U\rangle$, which is the same for all types of branes,
and a twisted component $|T;I\rangle$, which depends on the type of brane considered:
\begin{equation}
|\mathrm{D}3;I\rangle=\mathcal{N}\,|U\rangle+\mathcal{N}'\,|T;I\rangle
\label{BS0}
\end{equation}
where $\mathcal{N}$ and $\mathcal{N}'$ are appropriate 
normalization factors related to the brane tensions
(whose explicit expression is not relevant for our purposes).
This schematic structure holds of course both in the NS/NS and in the R/R sectors, which we now
discuss in turn, focusing on the fermionic twisted components.

\subsubsection{NS/NS sector}
The twisted component of the boundary state for a fractional
D3-brane of type $I$ is a sum of $(M-1)$ terms which refer to the $(M-1)$ twisted sectors of
the closed strings on the orbifold and whose coefficients have to be chosen in a specific way 
in order to have a consistent description of the D-brane. By this we mean that the cylinder amplitude 
between two such boundary states, once translated into the open string channel, must 
correctly reproduce the $\mathbb{Z}_M$-invariant one-loop annulus amplitude.  
In \cite{Billo:2000yb} a thorough analysis
of this issue was carried out in general, using the Cardy condition for the construction of
consistent boundary states in rational conformal field theories \cite{Cardy:1989ir}. 
Borrowing these results and adapting them to our case, we can write the twisted component 
of the boundary state for a D3-brane of type I in the NS/NS sector and its conjugate as follows:
\begin{equation}
\begin{aligned}
|T;I|\rangle_{\mathrm{NS}}
&=\sum_{\widehat{a}=1}^{{M-1}}\sin\Big(\frac{\pi\widehat{a}}{M}\Big)\,
\omega^{I\,\widehat{a}}\,|\,\widehat{a}\,\rangle\!\rangle_{\mathrm{NS}}~,\\
{}_{\mathrm{NS}}\langle T;I|
&=\sum_{\widehat{a}=1}^{{M-1}}\sin\Big(\frac{\pi\widehat{a}}{M}\Big)\,
\omega^{-I\,\widehat{a}}\,{}_{\mathrm{NS}}\langle\!\langle\,\widehat{a}\,|~.
\end{aligned}
\label{twistBSNS}
\end{equation}
Here, the sum runs over all twisted sectors, $\omega$ is the $M$-th root of unity as in (\ref{omega}) and $|\,\widehat{a}\,\rangle\!\rangle_{\mathrm{NS}}$ is the GSO projected
Ishibashi state for the twisted sector $\widehat{a}$. 
These Ishibashi states enforce the appropriate gluing
conditions between the left-moving and right-moving modes. For our purposes it is not necessary to write
the complete expression of these Ishibashi states, but it is enough to write the terms
which may have a non-zero overlap with the massless states of the closed string twisted sectors
discussed in Section~\ref{closedspectrum}. 

Let us suppose again that $M$ is odd. If $\widehat{a}=a\in[1,\frac{M-1}{2}]$, we have
\begin{equation}
|a\rangle\!\rangle_{\mathrm{NS}}=\Big(\ii\,\overbar{\Psi}^{\,2}_{-\frac{1}{2}+\nu_a}\,
\widetilde{\Psi^2}_{-\frac{1}{2}+\nu_a}-\ii\,{\Psi}^{3}_{-\frac{1}{2}+\nu_a}\,
\widetilde{\overbar{\Psi}^{3}}_{-\frac{1}{2}+\nu_a}
\Big)|\Omega_{a}\rangle_{(-1)} \,\,|\widetilde{\Omega}_{a}\rangle_{(-1)}+\cdots
\label{ishia}
\end{equation}
where the ellipses stand for terms involving a higher number of oscillators or massive fermionic 
modes. The relative minus sign in the brackets of (\ref{ishia}) 
is due to the fact that the complex direction 3 is transverse to the D3-brane
while the complex direction 2 is longitudinal.
If $\widehat{a}=(M-a)$, instead, we have
\begin{equation}
|M-a\rangle\!\rangle_{\mathrm{NS}}=\Big(\ii\,
{\Psi}^{2}_{-\frac{1}{2}+\nu_a}\,\widetilde{\overbar{\Psi}^2}_{-\frac{1}{2}+\nu_a}
-\ii\,\overbar{\Psi}^{\,3}_{-\frac{1}{2}+\nu_a}\,\widetilde{\Psi^3}_{-\frac{1}{2}+\nu_a}
\Big)|\Omega_{M-a}\rangle_{(-1)} \,\,|\widetilde{\Omega}_{M-a}\rangle_{(-1)}+\cdots~.
\label{ishiMa}
\end{equation}
The corresponding Ishibashi bra states are
\begin{equation}
\begin{aligned}
{}_{\mathrm{NS}}\langle\!\langle a |&=
{}_{(-1)}\langle\widetilde{\Omega}_{a}|\,\,{}_{(-1)}\langle\Omega_{a}|
\Big(\!-\ii\,\widetilde{\Psi^2}_{\frac{1}{2}-\nu_a}
\,\overbar{\Psi}^{2}_{\frac{1}{2}-\nu_a}+\ii\,
\widetilde{\overbar{\Psi}^3}_{\frac{1}{2}-\nu_a}\,
{\Psi}^{3}_{\frac{1}{2}-\nu_a}\Big)
+\cdots\\[2mm]
{}_{\mathrm{NS}}\langle\!\langle M-a|&={}_{(-1)}\langle \widetilde{\Omega}_{M-a}|\,\,
{}_{(-1)}\langle \Omega_{M-a}|
\Big(\!-\ii\,\widetilde{\overbar{\Psi}^2}_{\frac{1}{2}-\nu_a}\,
{\Psi}^{2}_{\frac{1}{2}-\nu_a}
+\ii\,\widetilde{\Psi^3}_{\frac{1}{2}-\nu_a}\,\overbar{\Psi}^{3}_{\frac{1}{2}-\nu_a}
\Big)+\cdots
\end{aligned}
\label{ishi1}
\end{equation}
where the conjugate vacuum states are normalized in such a way that
\begin{equation}
\begin{aligned}
{}_{(-1)}\langle \Omega_{a}|\Omega_{a}\rangle_{(-1)}&=1\quad\mbox{and}\quad
{}_{(-1)}\langle \Omega_{M-a}|\Omega_{M-a}\rangle_{(-1)}=1~,
\end{aligned}
\label{normvacuag}
\end{equation}
and similarly for the right-moving sectors.

In the presence of fractional D3-branes, the left and right moving parts 
of a twisted closed string have non-trivial correlation functions since
the closed string world-sheet has a boundary. In the boundary state formalism, this boundary 
is the unit circle on which the Ishibashi states enforce an identification between the left and 
the right movers of the closed strings. In particular for the massless vertex operators
of the twisted NS sector with label $a$ and twist parameter $\nu_a$ described in
Section~\ref{secn:NSNSa}, given any two points $w$ and $\bar{w}$ inside the unit disk 
$\mathbb{D}$ corresponding to a D3-brane of type $I$, we have
\begin{equation}
\begin{aligned}
\big\langle \mathcal{V}^{\alpha}_a(w)\,\widetilde{\mathcal{V}}^{\beta}_{a}(\bar{w})\big\rangle_I \,
&\equiv
\,{}_{\mathrm{NS}}\big\langle T;I|\,\mathcal{V}^{\alpha}_a(w)\,\widetilde{\mathcal{V}}^{\beta}_{a}(\bar{w})
|0\rangle|\widetilde{0}\big\rangle
=\frac{M^{\alpha\beta}_{I,a}}{(1-w\bar{w})^2}~,
\end{aligned}
\label{VabI}
\end{equation}
where the last step is a consequence of the conformal invariance which fixes the form of the
two-point function of conformal fields of weight 1 on $\mathbb{D}$.
The constant in the numerator can be obtained from the overlap between the twisted boundary
state and the states created by the vertex operators $\mathcal{V}^{\alpha}_a$ and $\widetilde{\mathcal{V}}^{\beta}_{a}$. For example, fixing $\alpha=1$ and $\beta=2$ and
referring to the explicit expressions in Table~\ref{tab:table5}, we have
\begin{align}
M^{12}_{I,a}&=\lim_{w\to 0}\,\lim_{\bar{w}\to 0}\,
{}_{\mathrm{NS}}\langle T;I|\,\mathcal{V}^{1}_a(w)\,\widetilde{\mathcal{V}}^{2}_{a}(\bar{w})
|0\rangle|\widetilde{0}\rangle\notag\\[1mm]
&={}_{\mathrm{NS}}\langle T;I| \,{\Psi}^{3}_{-\frac{1}{2}+\nu_a}\,
\widetilde{\overbar{\Psi}^3}_{-\frac{1}{2}+\nu_a}\,|\Omega_{a}\rangle_{(-1)}\, 
|\widetilde{\Omega}_{a}\rangle_{(-1)}\notag\\[1mm]
&=\sin(\pi\nu_a) \,\omega^{-Ia}\,{}_{\mathrm{NS}}\langle\!\langle a|\,{\Psi}^{3}_{-\frac{1}{2}+\nu_a}\,
\widetilde{\overbar{\Psi}^3}_{-\frac{1}{2}+\nu_a}\,|\Omega_{a}\rangle_{(-1)}\, |\widetilde{\Omega}_{a}\rangle_{(-1)}\notag\\[1mm]
&=\ii\,\sin(\pi\nu_a) \,\omega^{-Ia}~.
\label{m12}
\end{align}
Proceeding in a similar way, we find that $M^{21}_{I,a}$ is identical to (\ref{m12}), while
$M^{11}_{I,a}=M^{22}_{I,a}=0$, since in these cases the fermionic oscillators are unbalanced.
We can thus summarize this result by rewriting (\ref{VabI}) as
\begin{equation}
\big\langle \mathcal{V}^{\alpha}_a(w)\,\widetilde{\mathcal{V}}^{\beta}_{a}(\bar{w})
\big\rangle_{I} \,
=\frac{\ii\,\sin(\pi\nu_a) \,\omega^{-Ia}\,(\tau_1)^{\alpha\beta}}{(1-w\bar{w})^2}
\end{equation}
where $\tau_1$ is the first Pauli matrix.

We now map this disk two-point function onto the complex plane by using 
the Cayley transformation
\begin{equation}
w= \frac{z-\ii}{z+\ii}~,
\end{equation}
obtaining
\begin{equation}
\begin{aligned}
\big\langle \mathcal{V}^{\alpha}_a(z)\,\widetilde{\mathcal{V}}^{\beta}_{a}(\bar{z})\big\rangle_{I} \,
\,=\,\big\langle \mathcal{V}^{\alpha}_a(w)\,\widetilde{\mathcal{V}}^{\beta}_{a}(\bar{w})
\big\rangle_{I} \,\,\frac{dw}{dz} \,\frac{d\bar{w}}{d\bar{z}}
\,=\,
\frac{-\ii\,\sin(\pi\nu_a) \,\omega^{-Ia}\,(\tau^1)^{\alpha\beta}}{(z-\bar{z})^2}~.
\end{aligned}
\label{vvtilde}
\end{equation}
Thus, using the doubling trick, we are led to introduce the following reflection rule for right
moving vertex operators:
\begin{equation}
\widetilde{\mathcal{V}}^\beta_{a}(\bar{z})\longrightarrow
(R_{I,a})^\beta_{~\gamma}\,\mathcal{V}^\gamma_{M-a}(\bar{z})~,
\end{equation}
so that
\begin{equation}
\big\langle \mathcal{V}^{\alpha}_a(z)\,\widetilde{\mathcal{V}}^{\beta}_{a}(\bar{z})\big\rangle_{I}
\longrightarrow
(R_{I,a})^\beta_{~\gamma}\,
\big\langle \mathcal{V}^{\alpha}_a(z)\,{\mathcal{V}}^{\gamma}_{M-a}(\bar{z})\big\rangle=
(R_{I,a})^\beta_{~\gamma}\,\frac{(\epsilon^{-1})^{\alpha\gamma}}{(z-\bar{z})^2}
\end{equation}
where, in the last step, we used (\ref{vv}). Comparing with (\ref{vvtilde}) we find that the 
reflection matrix $R_{I,a}$ is given by 
\begin{equation}
\label{Rfora}
R_{I,a}=\ii\,\sin(\pi\nu_a) \,\omega^{-Ia}\, \tau_3
\end{equation}
where $\tau_3$ is the third Pauli matrix. Repeating the same calculations in the 
twisted sector labeled by $(M-a)$, we get
\begin{equation}
\label{RforM-a}
R_{I,M-a}=\ii\,\sin(\pi\nu_a) \,\omega^{Ia}\,\tau_3~.
\end{equation}
Notice that even though the oscillator structure of the boundary states in the sectors 
$a$ and $(M-a)$ is different, in the end the reflection matrices (\ref{Rfora}) and (\ref{RforM-a}) 
have the same form and can be simultaneously written as
\begin{equation}
R_{I,\widehat{a}}=\ii\,\sin\Big(\frac{\pi\widehat{a}}{M}\Big)\,\omega^{-I\,\widehat{a}}\,\tau_3
\label{Rhata}
\end{equation}
with $\widehat{a}=1,\ldots,M-1$.

\subsubsection{R/R sector}
\label{reflectionrulesRR}

The above analysis can be easily extended to the R/R sector where, in analogy with 
(\ref{twistBSNS}), the twisted components of the boundary state are given by
\begin{equation}
\begin{aligned}
|T;I|\rangle_{\mathrm{R}}
&=\sum_{\widehat{a}=1}^{{M-1}}\sin\Big(\frac{\pi\widehat{a}}{M}\Big)\,
\omega^{I\,\widehat{a}}\,|\,\widehat{a}\,\rangle\!\rangle_{\mathrm{R}}~,\\
{}_{\mathrm{R}}\langle T;I|
&=\sum_{\widehat{a}=1}^{{M-1}}\sin\Big(\frac{\pi\widehat{a}}{M}\Big)\,
\omega^{-I\,\widehat{a}}\,{}_{\mathrm{R}}\langle\!\langle\,\widehat{a}\,|~.
\end{aligned}
\label{twistRR}
\end{equation}
In writing the expressions for the GSO-projected Ishibashi states $|\,\widehat{a}\,\rangle\!\rangle_{\mathrm{R}}$ and their conjugates, we adopt the same picture assignments discussed
in Section~\ref{twistR}: the $(-\frac{1}{2},-\frac{3}{2})$-picture for the twisted sectors
labeled by $\widehat{a}=a\in [1,\frac{M-1}{2}]$, and the $(-\frac{3}{2},-\frac{1}{2})$-picture
for the sectors with $\widehat{a}=(M-a)$. Apart from this, the structure of these states is
similar to that of the twisted boundary states for D3-branes in the $\mathbb{Z}_2$ orbifold
obtained in \cite{Bertolini:2001gq} from the factorization of the one-loop open string partition
function, and already used in our companion paper \cite{Ashok:2020ekv}.
In particular, for $\widehat{a}=a$ we have
\begin{equation}
|a\rangle\!\rangle_{\mathrm{R}}= 
\big( C \Gamma_1 \Gamma_2)_{ A \dot B}
\,|A_a\rangle_{(-\frac{1}{2})}|\widetilde {\dot B }_a\rangle_{(-\frac{3}{2})}+\ldots
\end{equation}
where the ellipses stand for contributions from massive fermionic modes, the vacuum states
have been defined in (\ref{Rstates1}) and (\ref{Rstates2}), and $\Gamma_1$ and $\Gamma_2$
are the SO(6) Dirac matrices along the first two real longitudinal directions of the D3-branes.
Likewise, when $\widehat{a}=(M-a)$ we have
\begin{equation}
|M-a\rangle\!\rangle_{\mathrm{R}}= 
\big( C \Gamma_1 \Gamma_2)_{ \dot{A} B}
\,|\dot{A}_{M-a}\rangle_{(-\frac{3}{2})}|\widetilde {B}_{M-a}\rangle_{(-\frac{1}{2})}+\ldots~.
\end{equation}
The corresponding Ishibashi conjugate states are
\begin{equation}
\begin{aligned}
{}_{\mathrm{R}}\langle\!\langle a |&=\,\,
{}_{(-\frac{3}{2})} \langle\widetilde{ \dot A}_a|\,\,
{}_{(-\frac{1}{2})} \langle B_a| \,
\big (\Gamma_2 \Gamma_1 C^{-1}\big)^{ \dot A B}+\ldots
\\
{}_{\mathrm{R}} \langle\!\langle M-a| &=\,\,
{}_{(-\frac{1}{2})}\langle\widetilde{A}_{M-a}|\, \,
{}_{(-\frac{3}{2})}\langle \dot B_{M-a}| \,
\big (\Gamma_2 \Gamma_1 C^{-1}\big)^{A \dot B}+\ldots
\end{aligned}
\label{ishiRkdual}
\end{equation}
where the bra vacuum states are defined such that
\begin{equation}
{}_{(-\frac{1}{2})} \langle B_a|A_a\rangle_{(-\frac{1}{2})}=\delta_B^A
\quad\mbox{and}\quad
{}_{(-\frac{3}{2})} \langle \widetilde{\dot{B}}_{M-a}|\widetilde{\dot{A}}_{M-a}
\rangle_{(-\frac{3}{2})}=\delta_{\dot{B}}^{\dot{A}}
\label{overlapR}
\end{equation}
with analogous relations for the right-moving vacua\,%
\footnote{We remark that in (\ref{overlapR}) the superghost charges of the bra and ket states exactly soak up the background charge anomaly. For example the superghost charge of ${}_{(-\frac{1}{2})} \langle B_a|$ is $-\frac{3}{2}$, and that of $|A_a\rangle_{(-\frac{1}{2})}$ is $-\frac{1}{2}$.}.

We can now repeat the same steps followed in the NS sector to prove that the boundary state
enforces an identification between left-moving and right-moving vertex operators in the twisted R sector $a$
according to 
\begin{equation}
\widetilde{\mathcal{V}}^{\dot{B}}_{a}(\bar{z})\longrightarrow
(R_{I,a})^{\dot{B}}_{~\dot{C}}\,\mathcal{V}^{\dot{C}}_{M-a}(\bar{z})~,
\end{equation}
where the reflection matrix is the anti-chiral/anti-chiral block of
\begin{equation}
R_{I,a}=\sin(\pi\nu_a)\,\omega^{-Ia}\,\Gamma_1  \Gamma_2~.
\label{reflexRR}
\end{equation}
Similarly, in the twisted R sector labeled by $(M-a)$ the reflection matrix is the
chiral/chiral block of
\begin{equation}
R_{I,M-a}=\sin(\pi\nu_a)\,\omega^{Ia}\,\Gamma_1  \Gamma_2~.
\label{reflexRRMa}
\end{equation}
We can combine the last two formulas into
\begin{equation}
R_{I,\widehat{a}}=\sin\Big(\frac{\pi\widehat{a}}{M}\Big)\,\omega^{-I\widehat{a}}\,\Gamma_1  \Gamma_2
\end{equation}
with the understanding that one has to take the lower-right
and upper-left blocks for $\widehat{a}=a$ and $\widehat{a}=(M-a)$, respectively, as a consequence
of the picture assignments.

\subsection{Massless open string spectrum}
\label{openspectrum}

We now analyze the spectrum of massless open strings that live on a configuration made of stacks
of $n_I$ fractional D3-branes of type $I$ for $I=0,\ldots,M-1$, that engineer a theory with
gauge group $\mathrm{U}(n_0)\times\ldots\times\mathrm{U}(n_{M-1})$. We will restrict ourselves to listing the fields in the adjoint representation of U$(n_I)$ as these will be the only fields that are sourced by the background values given to the twisted closed string scalars. We tailor our notations and conventions to be as close as possible to those in \cite{Ashok:2020ekv}. 

In the familiar case of D3-branes in flat space, in the $(0)$-superghost picture the bosonic massless open string states are represented by vertex operators of the form\,%
\footnote{Here and in the following we always assume the operators to be normal ordered, unless this causes
ambiguities.}
\begin{equation}
\big(\ii\,\partial Z^i+\kappa \cdot \Psi\,\Psi^i\big)\, \rme^{\ii\,\kappa \cdot Z} ~.
\label{V}
\end{equation}
where
\begin{equation}
\kappa_i = \frac{k_{2i-1} + \ii\, k_{2i}}{\sqrt{2}}\quad \text{and}\quad \overbar{\kappa}_i = \frac{k_{2i-1} + \ii\, k_{2i}}{\sqrt{2}}~,
\label{kappai}
\end{equation}
with $k_{\mu}$ being the real momentum along the direction $x^{\mu}$ of the D3-brane world-volume.
In addition we denote the complex direction $1$ by the symbol $\parallel$ and the complex direction $2$ by the symbol $\perp$, since these directions are, respectively, 
longitudinal and perpendicular to the surface defect realized by the D3-brane configuration
on the orbifold. We also introduce the following convenient notation
\begin{equation}
\begin{aligned}
\kappa_\parallel \!\cdot\! Z_\parallel &
=\kappa_1\, \overbar{Z}^1+\overbar{\kappa}_1\,Z^1~,\quad
\kappa_\perp \!\cdot\! Z_\perp =\kappa_2\,\overbar{Z}^2+\overbar{\kappa}_2\,Z^2~,\\
\kappa_\parallel \!\cdot\! \Psi_\parallel &
=\kappa_1\, \overbar{\Psi}^1+\overbar{\kappa}_1\,\Psi^1~,\quad
\kappa_\perp \!\cdot\! \Psi_\perp =\kappa_2\,\overbar{\Psi}^2+\overbar{\kappa}_2\,\Psi^2~,
\end{aligned}
\end{equation}
so that
\begin{equation}
\kappa\cdot Z=
\kappa_\parallel \!\cdot\! Z_\parallel+\kappa_\perp \!\cdot\! Z_\perp
\end{equation}
and similarly for $\kappa\cdot \Psi$.
Clearly, the parallel terms $\kappa_\parallel \!\cdot\! Z_\parallel$ and 
$\kappa_\parallel \!\cdot\! \Psi_\parallel$ are invariant under the orbifold group $\mathbb{Z}_M$,
but the perpendicular terms are not, since
\begin{equation}
\label{gonZ}
g:~~ \begin{cases}
~~~
 \kappa_\perp \!\cdot\! Z_\perp ~~\longrightarrow &  g[\kappa_\perp \!\cdot\! Z_\perp ]
 =\omega^{-1}\kappa_2\,\overbar{Z}^2+\omega\,\overbar{\kappa}_2\,Z^2~, \\[1mm]
~~~
\kappa_\perp \!\cdot\! \Psi_\perp ~~\longrightarrow &  g[\kappa_\perp \!\cdot\! \Psi_\perp ]
 =\omega^{-1}\kappa_2\,\overbar{\Psi}^2+\omega\,\overbar{\kappa}_2\,\Psi^2~.
\end{cases}
\end{equation}
This in particular implies that in order to write the open string vertex operators for the fractional D3-branes
one cannot use the plane waves $\rme^{\ii\,\kappa_\perp\cdot Z_\perp}$ 
but instead decomposes these into functions that transform in the irreducible 
representations of $\mathbb{Z}_M$. These functions, which we denote by $\mathcal{E}_I$
with $I=0,\ldots,M-1$, are simply obtained by summing the
plane waves $\rme^{\ii\,\kappa_\perp \cdot Z_\perp}$ over the orbits of the group 
with coefficients chosen such that the combination transforms covariantly under the group action. So we are led to define:
\begin{align}
\label{em}
\mathcal{E}_I = 
\frac{1}{M}  \sum_{J=0}^{M-1} \omega^{-IJ} \, g^J\Big[\rme^{\ii \,\kappa_{\perp}\cdot  Z_{\perp}} \Big]= \frac{1}{M}  \sum_{J=0}^{M-1} \omega^{-IJ} \, \rme^{\ii\,(\omega^{-J}\kappa_2 \,\overbar{Z}^{2}
+\omega^{J} \overbar{\kappa}_2\, Z^{2})}~.
\end{align}
One can easily check that
\begin{align}
g\big[ \mathcal{E}_I\big]= \frac{1}{M}  \sum_{J=0}^{M-1} \omega^{-IJ} \, \rme^{\ii \,
(\omega^{-J-1}\kappa_2 \,\overbar{Z}^{2}
+\omega^{J+1} \overbar{\kappa}_2\, Z^{2})}
= \omega^I\,\mathcal{E}_I~,
\label{gonEz} 
\end{align}
which shows that 
$\mathcal{E}_I$ transforms in the $I$-th irreducible representation of $\mathbb{Z}_M$.
For $M=2$ and $\omega=-1$, the functions $\mathcal{E}_I$ are simply
\begin{equation}
\mathcal{E}_0 = \cos(\kappa_\perp\!\cdot\! Z_{\perp}) \quad \text{and}\quad 
\mathcal{E}_1 =  \ii\, \sin(\kappa_\perp\!\cdot \!Z_{\perp})~,
\end{equation}
which are exactly the two combinations used in the case of the $\mathbb{Z}_2$ orbifold in \cite{Ashok:2020ekv}. 

In a similar way, we have to break up the operators multiplying the plane wave in (\ref{V}) into various
pieces with definite charge $I$ under the orbifold action and form invariant combinations with
$\mathcal{E}_{M-I}$. In the orbifold theory, only such combinations
represent vertex operators describing physical fields on the world-volume of the fractional 
D3-brane.

Applying these considerations, we see that the gauge field $A_1$ along the parallel directions is described by the following vertex operator in the $(0)$-superghost picture:
\begin{equation}
\mathcal{V}_{A_1}=\Big[
\big(\ii\,\partial{Z}^1+\kappa_\parallel\!\cdot\! \Psi_\parallel\,{\Psi}^1\big)\,
\mathcal{E}_0+\kappa_{2}\,\overbar{\Psi}^{2}\,\Psi^1\, \mathcal{E}_1+ 
\overbar{\kappa}_{2}\,\Psi^{2}\,\Psi^1\, \mathcal{E}_{M-1}\Big]\,
\rme^{\ii\, \kappa_{\parallel}\cdot Z_{\parallel}} ~.
\label{VA1}
\end{equation}
Each term in square brackets is invariant under $\mathbb{Z}_M$. For instance, 
the terms $\partial \overbar{Z}^{1}$ and $\kappa_\parallel\!\cdot\! \Psi_\parallel\,
\overbar{\Psi}^1$, which are $\mathbb{Z}_M$ invariant, are multiplied with the invariant function
$\mathcal{E}_0$. Similarly the term $\kappa_{2}\,\overbar{\Psi}^{2}\,\Psi^1$,  
which gets a factor $\omega^{-1}$ under the orbifold action, 
is multiplied by $\mathcal{E}_1$ to make a $\mathbb{Z}_M$-invariant combination. 
Likewise, it is easy to see that the third term in (\ref{VA1}) is also ${\mathbb Z}_M$ invariant. 
The vertex operator for the complex conjugate field component $\overbar{A}_1$ is obtained by
simply replacing $\partial Z^1$ and $\Psi^1$ with $\partial \overbar{Z}^1$ and 
$\overbar{\Psi}^1$.

In a similar way we can write the vertex operators for the gauge field $A_2$ in the directions
transverse to the surface defect, which is 
\begin{align}
\mathcal{V}_{A_2}=\Big[\big(\ii\,\partial{Z}^2+\kappa_\parallel\!\cdot\! \Psi_\parallel\,{\Psi}^2
\big)\,\mathcal{E}_{M-1} + \kappa_{2}\,\overbar{\Psi}^{2}\,\Psi^2\, \mathcal{E}_0 \Big]\,
\rme^{\ii\, \kappa_{\parallel}\cdot Z_{\parallel}}~.
\label{VA2}
\end{align}
The vertex operator for $\overbar{A}_2$ can be obtained from the above expression
by replacing $\partial Z^2$ and $\Psi^2$ with $\partial \overbar{Z}^2$ and $\overbar{\Psi}^2$, and $\mathcal{E}_{M-1}$ with 
$\mathcal{E}_1$.

Finally, let us consider the scalar fields. On the fractional D3-brane world-volume there are 
three complex scalars that together with the gauge vector provide the bosonic content of 
the $\mathcal{N}=4$ vector multiplet. When the orbifold acts partially along the world-volume 
as in our case, all three complex scalars remain in the spectrum. Denoting them by $\Phi$ 
and $\Phi_r$ with $r=4,5$, they and their complex conjugates 
are described by the following $\mathbb{Z}_M$-invariant vertices:
\begin{equation}
\begin{aligned}
\mathcal{V}_{\Phi}&=\Big[
\big(\ii\,\partial{Z}^3+\kappa_\parallel\!\cdot\! \Psi_\parallel\,{\Psi}^3\big)\,
\mathcal{E}_{1} + \kappa_{2}\,\overbar{\Psi}^{2}\,\Psi^3\, \mathcal{E}_2
+\overbar{\kappa}_{2}\,\Psi^{2}\,\Psi^3\, \mathcal{E}_0\Big]
\,\rme^{\ii \,\kappa_{\parallel}\cdot Z_{\parallel}}~,\\
\mathcal{V}_{\overbar{\Phi}}&=\Big[
\big(\ii\,\partial\overbar{Z}^3+\kappa_\parallel\!\cdot\! \Psi_\parallel\,\overbar{\Psi}^3\big)\,
\mathcal{E}_{M-1} + \kappa_{2}\,\overbar{\Psi}^{2}\,\overbar{\Psi}^3\, \mathcal{E}_0
+\overbar{\kappa}_{2}\,\Psi^{2}\,\overbar{\Psi}^3\, \mathcal{E}_{M-2}\Big]
\,\rme^{\ii \,\kappa_{\parallel}\cdot Z_{\parallel}}~,
\label{VPhi} 
\end{aligned}
\end{equation}
and
\begin{equation}
\begin{aligned}
\mathcal{V}_{\Phi_r}&=\Big[
\big(\ii\,\partial{Z}^r+\kappa_\parallel\!\cdot\! \Psi_\parallel\,{\Psi}^r\big)\,
\mathcal{E}_0 +\kappa_{2}\,\overbar{\Psi}^{2}\,\Psi^r\,\mathcal{E}_1+
\overbar{\kappa}_{2}\,\Psi^{2}\,\Psi^r\, \mathcal{E}_{M-1}\Big]\,
\rme^{\ii\, \kappa_{\parallel}\cdot Z_{\parallel}}~,
\end{aligned}
\label{VPhii}
\end{equation}
with $\mathcal{V}_{\overbar{\Phi}_r}$ obtained by simply replacing ${\Psi}^r$ with
$\overbar{{\Psi}}^r$.

All these vertex operators have conformal dimension $1$ provided the corresponding fields are massless, {\it{i.e.}} if $\kappa\cdot\overbar{\kappa}=\frac{1}{2}k^2=0$.

\section{Open/closed correlators}
\label{diskcorr}

In this section we study the mixed amplitudes between the twisted closed string fields 
discussed in Section~\ref{closedspectrum} and the massless open string fields introduced 
in the previous section by calculating open/closed disk correlators 
(see \cite{Hashimoto:1996bf} for a review of scattering of strings off D-branes).
An example of such a mixed amplitude is shown in Figure~\ref{tadpole}, in which the closed string
field is the NS/NS scalar $b_{\alpha\beta}^{(\widehat{a})}$ in the twisted sector $\widehat{a}$. 

\begin{figure}[ht]
	\vspace{0.2cm}
	\begin{center}
		$\big\langle \mathcal{V}_{\text{open}} \big\rangle_{b^{(\widehat{a})}_{\alpha\beta};I}~~\equiv~~~$
		\parbox[c]{.28\textwidth}{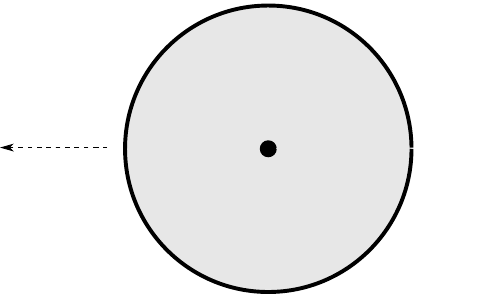}~.
	\end{center}
	\vspace{-0.2cm}
\caption{An example of a mixed open/closed string amplitude on a D3-brane of type $I$. The closed string vertex operator in the bulk represents the insertion of 
the twisted NS/NS scalar $b_{\alpha\beta}^{(\widehat{a})}$; the open string field on the boundary
is a generic massless excitation on the D3-brane which can couple to 
$b_{\alpha\beta}^{(\widehat{a})}$.
The result is a function of the open string momentum $\vec{k}_\perp$ 
along the two orbifolded directions of the D3-brane world-volume which are transverse to the surface defect.}
\label{tadpole}
\end{figure}
\noindent
The open/closed string amplitudes we consider correspond to disk diagrams with 
a closed string vertex inserted in the interior and an open string vertex inserted on the boundary. These diagrams are generically non-vanishing due to the D3-brane boundary conditions that enforce an identification between the left and right movers of the closed strings.

We now explain how to compute these mixed amplitudes starting from the NS/NS twisted
fields.

\subsection{Correlators with NS/NS twisted fields}
\label{NSNScorrelatorsexplicit}
Let us consider the scalar $b_{\alpha\beta}^{(\widehat{a})}$ in the NS/NS twisted sector 
$\widehat{a}$. Its coupling with a massless open string excitation on a D3-brane of type $I$
described by the vertex operator $\mathcal{V}_{\text{open}}$ is given 
by the following expression: 
\begin{equation}
\big\langle \mathcal{V}_{\text{open}} \big\rangle_{b^{(\widehat{a})}_{\alpha\beta};I}
=b^{(\widehat{a})}_{\alpha\beta}
\int\frac{dz\,d\bar{z}\,dx}{dV_{\text{proj}}}~\big\langle
\mathcal{V}_{\widehat{a}}^{\alpha}(z) \widetilde{\mathcal{V}}_{\widehat{a}}^{\beta}(\bar{z})\, 
\mathcal{V}_{\text{open}}(x)\big\rangle_I~,
\label{VaopenI}
\end{equation}
where
\begin{equation}
\label{proj_vol}
dV_{\text{proj}} = \frac{dz\,d\bar{z}\,dx}{(z-\bar{z})(\bar{z}-x)(x-z)}
\end{equation}
is the projective invariant volume element and the integrals are performed
on the string world-sheet. In particular the closed string insertion points $z$ and $\bar{z}$, are in the upper and lower half complex plane, respectively, while the open string insertion point $x$ is 
on the real axis.

Since we are interested in the couplings with constant background fields $b_{\alpha\beta}^{(\widehat{a})}$, the left and right vertex operators in (\ref{VaopenI}) 
are at zero momentum. The open string vertex, instead, has a non-vanishing momentum. 
Since the fractional brane is located at the orbifold fixed point $z_2=0$, translation invariance 
is broken in the complex direction 2. Therefore, 
the components $\kappa_2$ and $\overbar{\kappa}_2$ of the open string momentum
are arbitrary, while the components $\kappa_1$ and $\overbar{\kappa}_1$ are set to zero by 
momentum conservation in the parallel directions and the final amplitude will be proportional
to $\delta^{(2)}(\kappa_\parallel)$.

Using the reflection rule (\ref{Rhata}), the integrand of (\ref{VaopenI}) can be rewritten as
\begin{equation}
\begin{aligned}
\big\langle
\mathcal{V}_{\widehat{a}}^{\alpha}(z) \widetilde{\mathcal{V}}_{\widehat{a}}^{\beta}(\bar{z})\, 
\mathcal{V}_{\text{open}}(x)\big\rangle_I
&=\ii\,\sin\Big(\frac{\pi\widehat{a}}{M}\Big)\,
\omega^{-I\,\widehat{a}}\,(\tau_3)^{\beta}_{~\gamma}
\,\big\langle\mathcal{V}^{\alpha}_{\widehat{a}}(z)\,
\mathcal{V}^{\gamma}_{M-\widehat{a}}(\bar{z})\,\mathcal{V}_{\text{open}}(x)\big\rangle~.
\end{aligned}
\label{mixeda}
\end{equation}
Thus, the calculation is reduced to the evaluation of a three-point function of vertex operators
of conformal weight 1. The functional dependence on the word-sheet variables is fixed by
conformal invariance and exactly cancels that of the projective invariant volume (\ref{proj_vol})
so that in the end the result will be a constant that depends on the detailed structure of the vertex operators. 

There are, however, some features that can be described in generality, and are independent of
the specific components of $b_{\alpha\beta}^{(\widehat{a})}$ and of the particular open string
vertices that are considered. When we write the three-point functions in (\ref{mixeda}) as
products of correlators for each of the independent conformal fields, we easily
recognize that the superghost contribution is always given by
\begin{equation}
\big\langle\rme^{-\phi(z)}\,\rme^{-\phi(\bar{z})}\big\rangle = \frac{1}{z-\bar{z}}~.
\label{superghost}
\end{equation}
It is perhaps less obvious but it turns out that also the contribution arising from the bosonic string coordinates is the same for all amplitudes. Indeed, the only non-vanishing correlator involving the bosonic
coordinates along the parallel direction is
\begin{equation}
\big\langle \rme^{\ii\,\kappa_{\parallel}\cdot Z_\parallel}(x)\big\rangle = 
\delta^{(2)}(\kappa_\parallel)~,
\end{equation}
which enforces the anticipated momentum conservation 
for $\kappa_\parallel$, while the terms containing $\partial Z^1$ or $\partial \overbar{Z}^1$ always vanish inside the correlators and thus they can be ignored. 
As far as the perpendicular direction is concerned, we have to take into account the presence of the bosonic twist fields and the fact that the plane waves appear in the combinations $\mathcal{E}_I$ defined in (\ref{em}). Thus, one typically has to evaluate a correlator
of the form
\begin{equation}
\big\langle
\sigma_{\widehat{a}}(z)\, \sigma_{M-\widehat{a}}(\bar z)\,\mathcal{E}_I(x)
\big\rangle=\frac{1}{M}\sum_{J=0}^{M-1}
\omega^{-IJ}\,\big\langle
\sigma_{\widehat{a}}(z)\, \sigma_{M-\widehat{a}}(\bar z)\,\rme^{\ii\,\big(\omega^{-J}\kappa_2 \,\overbar{Z}^{2}(x)
+\omega^{J} \overbar{\kappa}_2\, Z^{2}(x)\big)}
\big\rangle~.
\label{ssze}
\end{equation}
For any value of $J$, the correlator in the sum is equal simply to 
$\langle \sigma_{\widehat{a}}(z)\, \sigma_{M-\widehat{a}}(\bar z)\rangle$, so that
\begin{equation}
\big\langle
\sigma_{\widehat{a}}(z)\, \sigma_{M-\widehat{a}}(\bar z)\,\mathcal{E}_I(x)
\big\rangle=\frac{1}{M}\Big(\sum_{J=0}^{M-1}\omega^{-IJ}\Big)
\big\langle
\sigma_{\widehat{a}}(z)\, \sigma_{M-\widehat{a}}(\bar z)\big\rangle
=\delta_{I,0}\,\big\langle
\sigma_{\widehat{a}}(z)\, \sigma_{M-\widehat{a}}(\bar z)\big\rangle~.
\label{ssze1}
\end{equation}
This means that in the open string vertex operators we can just focus on the terms proportional to
$\mathcal{E}_0$ and disregard the other terms, as they will not contribute. Furthermore, we can also neglect the terms involving $\partial Z^2$
or $\partial\overbar{Z}^2$, since they always give a vanishing contribution inside the correlators.
With this in mind, we can proceed to the explicit evaluation of the mixed amplitudes with
the twisted NS/NS scalars.

\subsubsection{Explicit computations}
We start by considering the correlator (\ref{mixeda}) with $\widehat{a}=a\in[1,\frac{M-1}{2}]$ and 
$\alpha=1$ and $\beta=2$, corresponding to the twisted field $b_{12}^{(a)}$. Applying the above
considerations, one realizes that this scalar does not couple to any open string field except
$A_2$ and $\overbar{A}_2$. Indeed, the terms of the vertex operators of $A_1$, $\Phi$,
$\Phi_r$ and their conjugates which contain $\mathcal{E}_0$ always contain other structures with unbalanced bosonic or fermionic fields, which therefore vanish inside the correlator. Let us then consider the coupling with $A_2$. In this case, inserting the explicit expressions of the vertex operators in (\ref{mixeda}), we have
\begin{equation}
\begin{aligned}
\big\langle
\mathcal{V}_{a}^{1}(z) \widetilde{\mathcal{V}}_{a}^{2}(\bar{z})\, 
\mathcal{V}_{A_2}(x)\big\rangle_I
&=-\ii\,\sin\pi\nu_a\,
\omega^{-Ia}\,\big\langle\mathcal{V}^{1}_{a}(z)\,
\mathcal{V}^{2}_{M-a}(\bar{z})\,\mathcal{V}_{A_2}(x)\big\rangle
\end{aligned}
\label{mixedA2}
\end{equation}
with
\begin{equation}
\begin{aligned}
\big\langle\mathcal{V}^{1}_{a}(z)\,
\mathcal{V}^{2}_{M-a}(\bar{z})\,\mathcal{V}_{A_2}(x)\big\rangle &=\kappa_2\,
\big\langle\rme^{-\phi(z)}\,\rme^{-\phi(\bar{z})}\big\rangle \, \big\langle \rme^{\ii\, \kappa_{\parallel}\cdot Z_{\parallel}(x)}\big\rangle\,
\big\langle
\sigma_{a}(z)\, \sigma_{M-a}(\bar z)\big\rangle\\[1mm]
&~~~\times\,\big\langle\!
:\!\Psi^3(z) s_{a}(z)\!:\,\,:\!\overbar{\Psi}^3(\bar z) s_{M-a}(\bar z)\!:\,\,:\!\overbar{\Psi}^2(x)\Psi^2(x)\!: \!\big\rangle~.
\end{aligned}
\label{ampl}
\end{equation}
The fermionic correlator in the second line above can be evaluated by factorizing it in the two independent directions 2 and 3 and using the bosonization method \cite{Kostelecky:1986xg}. 
In this way we have
\begin{equation}
\begin {aligned}
\big\langle\!
:\!\Psi^3(z) s_{a}(z)\!:\,\,:\!\overbar{\Psi}^3(\bar z) s_{M-a}(\bar z)\!:\,\,:\!\overbar{\Psi}^2(x)\Psi^2(x)\!: \!\big\rangle &= 
\big\langle s^2_{\nu_a}(z)\, s^2_{-\nu_a}(\bar z)
:\! \overbar{\Psi}^2 (x)\Psi^2 (x) \!:\!\big \rangle \\[2mm]
&~~~\times \big\langle\! :\!\Psi^3(z)s^3_{-\nu_a}(z)\!: \,\,:\!
\overbar{\Psi}^3(\bar z)s^3_{\nu_a}(\bar z)
\!:\!\big\rangle
\end{aligned}
\label{12a2psi}
\end{equation}
where\,%
\footnote{Here $\phi_2$ and $\phi_3$ denote the fields that bosonize the fermionic
systems in the complex directions 2 and 3.}
\begin{equation}
\begin {aligned}
\big\langle s^2_{\nu_a}(z)\, s^2_{-\nu_a}(\bar z)
:\! \overbar{\Psi}^2 (x)\Psi^2 (x) \!:\!\big \rangle 
&=\big\langle \, \rme^{\ii\,\nu_a\,\phi_2}(z) \, \rme^{-\ii\,\nu_a\,\phi_2}(\bar z)\,
(-\ii \,\partial \phi_2(x))\,\big\rangle\\[1mm]
&=\frac{-\nu_a}{(z-\bar z)^{\nu_a^2-1}(z-x)(\bar z-x)} ~,
\end{aligned}
\label{2a2psi2}
\end{equation}
and
\begin{equation}
\begin {aligned}
\big\langle\! :\!\Psi^3(z)s^3_{-\nu_a}(z)\!: \,\,:\!
\overbar{\Psi}^3(\bar z)s^3_{\nu_a}(\bar z)\!:\!\big\rangle &=
\big\langle \rme^{\ii\,(1-\nu_a)\phi_3}(z) \,\rme^{-\ii\,(1-\nu_a)\phi_3}(\bar z)\big\rangle
=\frac{1}{(z-\bar z)^{(1-\nu_a)^2}}~.
\end{aligned}
\label{12a2psi2}
\end{equation}
Combining everything together in (\ref{ampl}), we obtain
\begin{equation}
\big\langle\mathcal{V}^{1}_{a}(z)\,
\mathcal{V}^{2}_{M-a}(\bar{z})\,\mathcal{V}_{A_2}(x)\big\rangle=
\frac{\kappa_2\,\nu_a}{(z-\bar z)(\bar z-x)(x-z)}\,\delta^{(2)}(\kappa_\parallel)~.
\label{V12A2}
\end{equation}
Finally, inserting this into (\ref{mixedA2}) and (\ref{VaopenI}), we find 
that the coupling of $b_{12}^{(a)}$ with $A_2$ is
\begin{equation}
\big\langle \mathcal{V}_{A_2} \big\rangle_{b^{(a)}_{12};I}\,
=\,-\ii\, b^{(a)}_{12}\,\kappa_2\,\nu_a\,\sin\pi \nu_a \,\omega^{-Ia}\,\delta^{(2)}(\kappa_\parallel)~.
\label{Va12a2I}
\end{equation}
The same calculation shows that $b_{12}^{(a)}$ also couples to $\overbar{A}_2$ and the result
is simply obtained by replacing $\kappa_2$ with $-\overbar{\kappa}_2$ in the above
expression.

We can similarly repeat the analysis for the other components $b^{(a)}_{\alpha\beta}$.
For example, taking $b^{(a)}_{21}$ we find that its only non-vanishing coupling is
\begin{equation}
\big\langle \mathcal{V}_{A_2} \big\rangle_{b^{(a)}_{21};I}\,
=\,\ii\, b^{(a)}_{21}\,\kappa_2\,(1-\nu_a)\,\sin\pi \nu_a \,\omega^{-Ia}
\,\delta^{(2)}(\kappa_\parallel)~,
\label{Va21a2I}
\end{equation}
with a similar result for $\overbar{A}_2$ in which $\kappa_2$ is replaced 
with $-\overbar{\kappa}_2$.
The diagonal components $b^{(a)}_{11}$ and $b^{(a)}_{22}$, instead, only couple to the
complex scalars $\Phi$ and $\overbar{\Phi}$ according to
\begin{equation}
\begin{aligned}
\big\langle \mathcal{V}_{\Phi} \big\rangle_{b^{(a)}_{22};I}\,
&=-\ii\, b^{(a)}_{22}\,\overbar{\kappa}_2\,\sin\pi \nu_a \,\omega^{-Ia}
\,\delta^{(2)}(\kappa_\parallel)~,\\
\text{and}\quad \big\langle \mathcal{V}_{\overbar{\Phi}} \big\rangle_{b^{(a)}_{11};I}\,
&=\phantom{-}\,\ii\, b^{(a)}_{11}\,\kappa_2\,\sin\pi \nu_a \,\omega^{-Ia}
\,\delta^{(2)}(\kappa_\parallel)~.
\end{aligned}
\label{Phib11b22}
\end{equation}

It is equally straightforward to compute the open/closed string correlators in the twisted sectors
with $\widehat{a}=(M-a)$. In this case, we find again that the off-diagonal components 
$b_{12}^{(M-a)}$ and $b_{21}^{(M-a)}$ only interact with $A_2$ and $\overbar{A}_2$, and
that the couplings with $A_2$ are
\begin{equation}
\begin{aligned}
\big\langle \mathcal{V}_{A_2} \big\rangle_{b^{(M-a)}_{12};I}\,
&=-\ii\, b^{(M-a)}_{12}\,\kappa_2\,(1-\nu_a) \sin\pi\nu_a \,\omega^{Ia}
\,\delta^{(2)}(\kappa_\parallel)~,\\
\text{and}\quad \big\langle \mathcal{V}_{A_2} \big\rangle_{b^{(M-a)}_{21};I}\,
&=\phantom{-}\,\ii\, b^{(M-a)}_{21}\,\kappa_2\,\nu_a\,\sin\pi \nu_a \,\omega^{Ia}
\,\delta^{(2)}(\kappa_\parallel)~,
\end{aligned}
\label{b1221Ma}
\end{equation}
while those with $\overbar{A}_2$ follow by replacing $\kappa_2$ with $-\overbar{\kappa}_2$
in the above expressions. The diagonal components $b_{11}^{(M-a)}$ and $b_{21}^{(M-a)}$
interact instead with $\Phi$ and $\overbar{\Phi}$ with the following couplings:
\begin{equation}
\begin{aligned}
\big\langle \mathcal{V}_{\Phi} \big\rangle_{b^{(M-a)}_{22};I}\,
&=-\ii\, b^{(M-a)}_{22}\,\overbar{\kappa}_2\,\sin\pi \nu_a \,\omega^{Ia}
\,\delta^{(2)}(\kappa_\parallel)~,\\
\text{and}\quad \big\langle \mathcal{V}_{\overbar{\Phi}} \big\rangle_{b^{(M-a)}_{11};I}\,
&=\phantom{-}\,\ii\, b^{(M-a)}_{11}\,\kappa_2\,\sin\pi \nu_a \,\omega^{Ia}
\,\delta^{(2)}(\kappa_\parallel)~.
\end{aligned}
\label{Phib11b22M}
\end{equation}
As a consistency check of our results, we observe that the formulas (\ref{b1221Ma}) and
(\ref{Phib11b22M}) can be obtained from (\ref{Va12a2I}), (\ref{Va21a2I}) and
(\ref{Phib11b22}) by simply replacing everywhere $a$ with $(M-a)$. Thus, despite the fact that 
the fermionic approach we have used introduces differences in the explicit expressions 
for the twisted sector vertex operators, in the end, all sectors are treated on an equal footing.

\subsubsection{Results}
We are finally in a position to write down the complete expression for the open string
fields emitted by a fractional D3-brane of type $I$ in the presence of background values for 
the scalars of the NS/NS  twisted sectors. This is given by summing over all components of 
$b_{\alpha\beta}^{(\widehat{a})}$ and over all twisted sectors:
\begin{align}
\big\langle \mathcal{V}_{\text{open}} \big\rangle_{I} =
\sum_{\widehat{a}=1}^{M-1} 
\sum_{\alpha, \beta=1}^2 \big\langle \mathcal{V}_{\text{open}} \big\rangle_{b^{(\widehat{a})}_{\alpha\beta};I}~.
 \label{VopenaIdefn}
\end{align}
As we have seen, the components of the gauge field along the parallel direction 1 and
the complex scalars $\Phi_r$ do not couple to any NS/NS twisted field, while we have a non-vanishing
source for $A_2$, $\Phi$ and their complex conjugates. For $A_2$ the above formula gives
\begin{equation}
\begin{aligned}
\big\langle \mathcal{V}_{A_2} \big\rangle_{I} &=-
\ii\,\kappa_2\, \sum_{a=1}^{\frac{M-1}{2}}
\sin\pi \nu_a \, \Big[\nu_a\,\omega^{-Ia}\,b^{(a)}_{12} - (1-\nu_a)\,\omega^{-Ia}\,b^{(a)}_{21}
\\&\hspace{3cm} -\nu_a\,\,\omega^{Ia}\,b^{(M-a)}_{21} +(1-\nu_a)\,\omega^{Ia}\, b^{(M-a)}_{12}
\Big]\,\delta^{(2)}(\kappa_\parallel)~.
\end{aligned}
\label{a2tot}
\end{equation}
Taking into account the relations (\ref{starNSexp}), it is easy to realize that the quantity in square brackets is purely imaginary. A similar result holds for $\overbar{A}_2$ with $\kappa_2$ replaced
by $-\overbar{\kappa}_2$. 

For the complex scalars $\Phi$ and $\overbar{\Phi}$ we have instead
\begin{equation}
\begin{aligned}
\big\langle \mathcal{V}_{\Phi} \big\rangle_{I} &=-\ii\,\overbar{\kappa}_2\,
\sum_{a=1}^{\frac{M-1}{2}}
\sin\pi \nu_a \, \Big[\omega^{-Ia}\,b_{22}^{(a)}+\omega^{Ia}\,b_{22}^{(M-a)}\Big]
\,\delta^{(2)}(\kappa_\parallel)~,\\
\big\langle \mathcal{V}_{\overbar{\Phi}} \big\rangle_{I} &=\phantom{-}\,\ii\,\kappa_2\,
\sum_{a=1}^{\frac{M-1}{2}}
\sin\pi \nu_a \, \Big[\omega^{-Ia}\,b_{11}^{(a)}+\omega^{Ia}\,b_{11}^{(M-a)}\Big]
\,\delta^{(2)}(\kappa_\parallel)~.
\end{aligned}
\label{VPhibarPhi}
\end{equation}

\subsection{Correlators with R/R twisted fields}
\label{worldsheetcorr}
We now turn to the calculation of the interactions between the massless open string fields and the twisted R/R potentials. For definiteness we only consider non-vanishing background values for the scalars $\mathcal{C}^{(a)}$
and $\mathcal{C}^{(M-a)}$, since they are the only ones that turn out to be 
relevant for the description of the continuous parameters of surface defects. 
Thus, the closed string vertex operators we consider are
\begin{equation}
\mathcal{C}^{(a)}\,C_{A\dot{B}}\,\cV^{A}_a(z)\,\widetilde{\cV}^{\dot{B}}_a(\bar{z})\quad
\mbox{and}\quad
\mathcal{C}^{(M-a)}\,C_{\dot{A}B}\,\cV^{\dot{A}}_{M-a}(z)\,\widetilde{\cV}^{B}_{M-a}(\bar{z})~.
\end{equation}
By inspecting the fermionic structure of these vertex operators and comparing it with that 
of the open string vertices, one realizes that only the longitudinal component of the gauge field $A_1$ and its conjugate $\overbar{A}_1$ can 
have a non-vanishing coupling.

Let us start by considering the interaction between $A_1$ and $\mathcal{C}^{(a)}$. 
This is given by
\begin{equation}
\big\langle \mathcal{V}_{A_1} \big\rangle_{\mathcal{C}^{(a)}, I}
=\mathcal{C}^{(a)}\,C_{A\dot{B}}\,\int\frac{dz\,d\bar{z}\,dx}{dV_{\text{proj}}}~\big\langle
\cV^{A}_a(z)\,\widetilde{\cV}^{\dot{B}}_a(\bar{z})\,\mathcal{V}_{A_1}(x)\big\rangle_I~,
\label{VCAI}
\end{equation}
where the projective invariant volume element is defined in (\ref{proj_vol}). 
Using the reflection rules (\ref{reflexRR}) for the R/R fields, the integrand of 
(\ref{VCAI}) becomes
\begin{equation}
\big\langle\mathcal{V}^{A}_a(z)\,
\widetilde{\mathcal{V}}^{\dot B}_a(\bar{z})\,\mathcal{V}_{A_1}(x)\big\rangle_I =\sin(\pi\nu_a)\,\omega^{-Ia}\, (\Gamma_1\Gamma_2)^{\dot{B}}_{~\dot C}
\,\big\langle\mathcal{V}^{A}_a(z)\,
\mathcal{V}^{\dot C}_{M-a}(\bar{z})\,\mathcal{V}_{A_1}(x)\big\rangle~.
\label{mixedaRR}
\end{equation}
Using the explicit form of the vertex operators given in (\ref{vertexR1}), (\ref{vertexR311})
and (\ref{VA1}), and taking into account the points discussed at the beginning of this section, 
the above correlator can be written as follows:
\begin{equation}
\begin{aligned}
\big\langle\mathcal{V}^{A}_a(z)\,
\mathcal{V}^{\dot C}_{M-a}(\bar{z})\,\mathcal{V}_{A_1}(x)\big\rangle &=\kappa_1
\big\langle\rme^{-\frac{1}{2}\phi(z)}\,\rme^{-\frac{3}{2}\phi(\bar{z})}\big\rangle \, 
\big\langle \rme^{\ii \,\kappa_{\parallel}\cdot Z_{\parallel}}\big \rangle \,
\big\langle \sigma_{a}(z)\, \sigma_{M-a}(\bar z)\big\rangle\\[1mm]
&\hspace{2cm} \times \,\big\langle r_a(z) \,r_{M-a}(\bar z) \big\rangle\,
\big\langle S^A(z)\, S^{\dot C}(\bar z) \, :\! \overbar{\Psi}^1 \Psi^1\!:(x) \big\rangle~.
\end{aligned}
\label{ACdota1}
\end{equation}
Each factor in this expression can be easily computed using standard conformal field theory methods. The new ingredients with respect to the calculations in the NS/NS sectors are the following
two-point functions:
\begin{equation}
\begin{aligned}
\big\langle\rme^{-\frac{1}{2}\phi(z)}\,\rme^{-\frac{3}{2}\phi(\bar{z})}\big\rangle &=
\frac{1}{(z-\bar{z})^{\frac{3}{4}}}~,\\
\big\langle r_a(z) \,r_{M-a}(\bar z) \big\rangle &=
\frac{1}{(z-\bar{z})^{\frac{1}{2}-2\nu_a(1-\nu_a)}}~,\\
\text{and}\quad \big\langle S^A(z)\, S^{\dot C}(\bar z) \, :\! \overbar{\Psi}^1 \Psi^1\!:(x) \big\rangle &=
\frac{\ii}{2}\,\frac{(\Gamma_1\Gamma_2C^{-1})^{A\dot{C}}}
{(z-\bar{z})^{-\frac{1}{4}}(z-x)(\bar{z}-x)}~.
\end{aligned}
\end{equation}
Putting everything together, we have
\begin{equation}
\big\langle\mathcal{V}^{A}_a(z)\,
\mathcal{V}^{\dot C}_{M-a}(\bar{z})\,\mathcal{V}_{A_1}(x)\big\rangle=
-\frac{\ii}{2}\,\frac{(\Gamma_1\Gamma_2C^{-1})^{A\dot{C}}}{(z-\bar{z})(\bar{z}-x)(x-z)}\,
\,\delta^{(2)}(\kappa_\parallel)~.
\end{equation}
Inserting this into (\ref{mixedaRR}) and (\ref{VCAI}), 
and performing the $\Gamma$-matrix algebra, we finally obtain
\begin{equation}
\big\langle \mathcal{V}_{A_1} \big\rangle_{\mathcal{C}^{(a)}, I}
=-2\ii\,\kappa_1\,\sin \pi\nu_a \,\omega^{-Ia}\,\mathcal{C}^{(a)}\,\delta^{(2)}(\kappa_\parallel)~.
\label{VCafin}
\end{equation}
In a very similar way we find
\begin{equation}
\big\langle \mathcal{V}_{A_1} \big\rangle_{\mathcal{C}^{(M-a)}, I}
=-2\ii\,\kappa_1\,\sin \pi\nu_a \,\omega^{Ia}\,\mathcal{C}^{(M-a)}\,\delta^{(2)}(\kappa_\parallel)~.
\label{VCMfin}
\end{equation}
Thus, the full amplitude becomes
\begin{equation}
\big\langle \mathcal{V}_{A_1} \big\rangle_{I}=
-2\ii\,\kappa_1\,\sum_{a=1}^{\frac{M-1}{2}} 
\Big[\sin \pi\nu_a \big(\omega^{-Ia}\,\mathcal{C}^{(a)}+
\omega^{Ia}\,\mathcal{C}^{(M-a)}\big)\Big]
\,\delta^{(2)}(\kappa_\parallel)~.
\label{VA1Cafin}
\end{equation}
Taking into account that $\mathcal{C}^{(M-a)}=\mathcal{C}^{(a)\,\star}$, as it follows from
(\ref{starR}), we see that the expression inside the square brackets is real.

\section{Continuous parameters of surface defects}
\label{profilesfromFT}

We are now ready to identify the twisted closed string background that leads to a monodromy 
surface defect in the gauge theory on the world-volume of the fractional D3-branes.
It is convenient to decompose the twisted fields of the NS/NS sectors into irreducible representations
of the unbroken SU$(2)_+$ symmetry group of the orbifolded space (see the discussion in Section \ref{closedspectrum}). In each twisted sector $\widehat{a}$, this can be done by writing 
\begin{equation}
b_{\alpha\beta}^{(\widehat{a})}=\ii\,b_{\text{s}}^{(\widehat{a})}\,\epsilon_{\alpha\beta}+
b_+^{(\widehat{a})}\,(\epsilon\tau_+)_{\alpha\beta}+
b_-^{(\widehat{a})}\,(\epsilon\tau_-)_{\alpha\beta}+
b_3^{(\widehat{a})}\,(\epsilon\tau_3)_{\alpha\beta}
\end{equation}
where $\epsilon$ is defined in (\ref{epsilon}) and $\tau_{\pm}=(\tau_1\pm\,\ii\tau_2)/2$.
In the $M=2$ case studied in \cite{Ashok:2020ekv} it was found that only the singlet 
component $b_{\text{s}}^{(\widehat{a})}$ (which we denoted $b$ in that reference) acted as a source for the gauge field. This can also be
seen from (\ref{a2tot}) by setting $\nu_1=\frac{1}{2}$ and $\omega=-1$ for the only twisted
sector that is present when $M=2$. For the general $M>2$ case, however, we see that the gauge field couples to both the scalars $b^{(\widehat{a})}_{\text s}$ and $b^{(\widehat{a})}_{3}$. Since we wish to have a uniform description of surface defects for all values of $M$, in what follows, we will set $b^{(\widehat{a})}_{3}=0$ and only turn on the background value for $b^{(\widehat{a})}_{\text s}$.
Furthermore, we also turn on the doublet components
$b^{(\widehat{a})}_{\pm}$ which source the scalar fields $\Phi$ and $\overbar{\Phi}$.
This means that, in terms of the initial fields $b_{\alpha\beta}^{(\widehat{a})}$, our background
reads
\begin{equation}
\begin{aligned}
b_{12}^{(\widehat{a})}&=-b_{21}^{(\widehat{a})}=-\ii\,b^{(\widehat{a})}_{\text s}~,\\
b_{22}^{(\widehat{a})}&=b_{+}^{(\widehat{a})}~,\quad
b_{11}^{(\widehat{a})}=-b_{-}^{(\widehat{a})}~,
\end{aligned}
\label{bs}
\end{equation}
with $(b^{(\widehat{a})}_{\text s}\big)^*=b^{(M-\widehat{a})}_{\text s}$ and
$(b^{(\widehat{a})}_{+}\big)^*=b^{(M-\widehat{a})}_{-}$ for all twisted sectors,
as follows from the relations (\ref{starNSexp}). 

Inserting these background values in (\ref{a2tot}) and (\ref{VPhibarPhi}), we have
\begin{equation}
\big\langle \mathcal{V}_{A_2} \big\rangle_{I} =-\kappa_2\,b_I\,\delta^{(2)}(\kappa_\parallel)~,
\label{A2final}
\end{equation}
and
\begin{equation}
\big\langle \mathcal{V}_{\Phi} \big\rangle_{I} =-\ii\,\overbar{\kappa}_2\,b_I^+\,
\delta^{(2)}(\kappa_\parallel)~,\quad
\big\langle \mathcal{V}_{\overbar{\Phi}} \big\rangle_{I} =-\ii\,\kappa_2\,b_I^-\,
\delta^{(2)}(\kappa_\parallel)~,
\label{Phifinal}
\end{equation}
where we have defined the combinations
\begin{equation}
b_I=\sum_{a=1}^{\frac{M-1}{2}}
\sin\pi \nu_a \, \Big[\omega^{-Ia}\,b^{(a)}_{\text s} +\omega^{Ia}\,b^{(M-a)}_{\text s} 
\Big]=\sum_{\widehat{a}=1}^{M-1}\sin\Big(\frac{\pi\widehat{a}}{M}\Big)\,
\omega^{-I\widehat{a}}\,b^{(\widehat{a})}_{\text s} ~,
\label{bI}
\end{equation}
and
\begin{equation}
b_I^{\pm}=\sum_{a=1}^{\frac{M-1}{2}}
\sin\pi \nu_a \, \Big[\omega^{-Ia}\,b^{(a)}_{\pm} +\omega^{Ia}\,b^{(M-a)}_{\pm} 
\Big]=\sum_{\widehat{a}=1}^{M-1}\sin\Big(\frac{\pi\widehat{a}}{M}\Big)\,
\omega^{-I\widehat{a}}\,b^{(\widehat{a})}_{\pm} ~.
\label{bIpm}
\end{equation}
Notice that $b_I$ is real, while $(b_I^+)^*=b_I^-$.
It is interesting to note that a similar change of basis for profiles of closed string fields between the fractional branes (labelled by irreducible representations) and the twisted sectors (labelled by conjugacy classes) has been observed previously for fractional branes at orbifolds in \cite{Billo:2001vg}.

As explained in detail in \cite{Ashok:2020ekv}, these amplitudes are interpreted as a source
for the corresponding open string field (see also Figure~\ref{tadpole}), 
whose profile in configuration space is obtained
by taking the Fourier transform, after attaching the massless
propagator along the D3-brane world-volume:
\begin{equation}
\frac{1}{k^2}=\frac{1}{2\big(|\kappa_\parallel|^2+|\kappa_\perp|^2\big)}~.
\end{equation}
For example, for the gauge field $A_2$ we have
\begin{equation}
A_{2;I} = \mathcal{FT}\bigg[\frac{\big\langle \mathcal{V}_{A_2} \big\rangle_{I}}{k^2}\bigg]~.
\label{A2FT}
\end{equation}
In Appendix \ref{sec:orbZM} we show how to organize the calculation of this Fourier transform in terms of the generalized plane-waves $\mathcal{E}_I$ that transform covariantly with charge
$I$ under the orbifold group. Applying these methods to the present case, we see that since
the source (\ref{A2final}) is proportional to $\kappa_2$, which has charge $(-1)$, only the term 
proportional to $\mathcal{E}_1$ remains so that (\ref{A2FT}) becomes
\begin{equation}
\begin{aligned}
A_{2;I} &= \int \frac{d^2\kappa_\parallel d^2\kappa_\perp}{(2\pi)^2} \frac{\big\langle \cV_{A_2} \big\rangle_I}{2(\kappa_\parallel^2+\kappa^2_\perp)}\, \rme^{\ii\,\kappa_\parallel\cdot
z_\parallel}\,\mathcal{E}_1\\
&=-b_I\,\frac{1}{M}\sum_{J=0}^{M-1} \omega^{-J}\!\int \frac{d^2\kappa_\perp}{(2\pi)^2}\, \frac{\kappa_2}{2|\kappa_\perp|^2} \,
\rme^{\ii\, (\omega^{-J}\kappa_2\,\bar{z}_2+\omega^J\,\overbar{\kappa}_2\, z_2)}
\,=\,-\frac{\ii\,b_I}{4\pi \bar{z}_2} ~,
\end{aligned}
\label{A2profile}
\end{equation}
where the last equality is a consequence of the fact that all $M$ terms in the sum are actually all equal to
each other and equal to ${\ii}/{(4\pi \bar{z}_2)}$.

Combining this result with the one for the complex conjugate component $\overbar{A}_2$, 
we find that the gauge field on the $I$-th fractional D3-brane has the following profile:
\begin{equation}
{\bf A}_I= A\cdot dx= A_{2;I} \,d\bar{z}_2 + \overbar{A}_{2;I }\,dz_2  
= -\frac{\ii\, b_I}{4\pi}\Big(\frac{d\bar{z}_2}{\bar{z}_2} - \frac{dz_2}{z_2}\Big) 
= -\frac{b_I}{2\pi}\, d\theta~,
\end{equation}
where $\theta$ is the polar angle in the  ${\mathbb C}_{(2)}$-plane transverse to the surface
defect.

The only other open string field that has a non-vanishing profile in the twisted
NS/NS background we have chosen is the complex scalar $\Phi$. 
The analogous calculation takes the following form:
\begin{equation}
\begin{aligned}
\Phi_{I} &=\mathcal{FT}\bigg[\frac{\big\langle \mathcal{V}_{\Phi} \big\rangle_{I}}{k^2}\bigg]= 
\int \frac{d^2\kappa_\parallel d^2\kappa_\perp}{(2\pi)^2} \frac{\big\langle \cV_{\Phi}
 \big\rangle_I}{2(\kappa_\parallel^2+\kappa^2_\perp)}\, \rme^{\ii\,\kappa_\parallel\cdot
z_\parallel}\,\mathcal{E}_{M-1}\\
&=-\ii\,b_I^+\,\frac{1}{M}\sum_{J=0}^{M-1} \omega^{J}\!\int \frac{d^2\kappa_\perp}{(2\pi)^2}\, \frac{\overbar{\kappa}_2}{2|\kappa_\perp|^2} \,
\rme^{\ii\, (\omega^{J}\kappa_2\,\bar{z}_2+\omega^J\,\overbar{\kappa}_2\, z_2)}
\,=\,\frac{b_I^+}{4\pi z_2}~.
\end{aligned}
\label{Phiprofile0}
\end{equation}
If we now consider a general configuration with $n_I$ fractional D3-branes of type $I$ for all values
of $I$, as in the KT proposal \cite{Kanno:2011fw}, we obtain the following profiles:
\begin{equation}
\mathbf{A} =  -\frac{d\theta}{2\pi}\begin{pmatrix}
		b_0\,\mathbb{1}_{n_0}&0&\cdots&0\\
		0&b_1\,\mathbb{1}_{n_1}&\cdots&0\\
		\vdots&\vdots&\ddots&\vdots\\
		0&0&\cdots&b_{M-1}\,\mathbb{1}_{n_{M-1}}
		\end{pmatrix} ~, 
		\label{Aprofile}
\end{equation}
and
\begin{equation}
\mathbf{\Phi} =  \frac{1}{4\pi\, z_2}\begin{pmatrix}
		b_{0}^+\,\mathbb{1}_{n_0}&0&\cdots&0\\
		0&b_{1}^+\,\mathbb{1}_{n_1}&\cdots&0\\
		\vdots&\vdots&\ddots&\vdots\\
		0&0&\cdots&b_{M-1}^+
		\end{pmatrix}~.
		\label{Phiprofile}
\end{equation}
These are precisely the profiles of a GW surface defect in the $\mathcal{N}=4$
theory corresponding to the breaking of U($N$) group to the Levi subgroup 
$\mathrm{U}(n_0)\times\ldots \times \mathrm{U}(n_{M-1})$, provided the continuous parameters 
$(\alpha_I, \beta_I, \gamma_I)$ that conventionally parametrize the singular profiles near the defect are related to the background values of the NS/NS twisted scalars as follows: 
\begin{equation}
\alpha_I = -\frac{b_I}{2\pi}~, \qquad \beta_I =\frac{\text{Re}(b_{I}^+)}{2\pi}~, 
\qquad \gamma_I = \frac{\text{Im}(b_{I}^+)}{2\pi}~.
\label{abgare}
\end{equation}
If the original gauge group is SU($N$), the corresponding field profiles are obtained by removing
the overall trace from each of the above expressions.

We now turn to discussing the coupling of the open string fields with 
the twisted scalars in the R/R sector. As we have seen in Section~\ref{worldsheetcorr}, we 
only need to consider the coupling with the longitudinal component $A_1$ of the gauge field.
This is given in (\ref{VA1Cafin}), which we rewrite as
\begin{equation}
\big\langle \mathcal{V}_{A_1} \big\rangle_{I}=
-2\ii\,\kappa_1c_I\,\delta^{(2)}(\kappa_\parallel)
\label{VA1Cafinal}
\end{equation}
where 
\begin{equation}
c_I=\sum_{a=1}^{\frac{M-1}{2}} 
\sin\pi \nu_a \, \Big[\omega^{-Ia}\,\mathcal{C}^{(a)} +\omega^{Ia}\,\mathcal{C}^{(M-a)}
\Big]=\sum_{\widehat{a}=1}^{M-1}\sin\Big(\frac{\pi\widehat{a}}{M}\Big)\,
\omega^{-I\widehat{a}}\,\mathcal{C}^{(\widehat{a})}~.
\label{cI}
\end{equation}
This real quantity is the R/R counterpart of $b_I$ defined in (\ref{bI}) for the NS/NS
sectors.

At face value, the coupling (\ref{VA1Cafinal}) is vanishing because of the $\delta$-function.
However, as was explained in the $\mathbb{Z}_2$ in \cite{Ashok:2020ekv}, if we multiply
this amplitude and its complex conjugate with the corresponding gauge field polarizations, 
the resulting sum can be interpreted as an interaction term between the R/R scalars 
and the longitudinal components of the gauge field strength. Indeed,
\begin{equation}
\overbar{A}_{1, I}\,\big\langle \mathcal{V}_{A_1} \big\rangle_I+ 
A_{1,I}\,\big\langle \mathcal{V}_{\overbar{A}_{1}} \big\rangle_I 
= -2\, \ii\, c_I \big(\overbar{\kappa}_1 A_1 - \kappa_1 \overline {A}_{1}\big)\, 
\delta^2(\kappa_\parallel) = 2\, \ii\, c_I\, \widetilde{F}_I\, \delta^2(\kappa_\parallel)~,
\end{equation}
where $\widetilde{F}_I$ is the gauge field strength on the $I$th fractional brane (along the defect), in momentum space.
Performing the Fourier transform, this expression becomes an effective interaction term
localized on the surface defect:
\begin{equation}
\frac{\ii\,c_I}{2\pi}\int\!d^2z_\parallel\,F_I~,
\end{equation}
where $F_I$ is the gauge field strength in configuration space, on the $I$th fractional brane. If this has a non-trivial first 
Chern class, then this effective interaction can be understood as the 2$d$ topological 
$\theta$-term that can be included in the path integral definition of the theory with surface defect. 
When a generic configuration with $n_I$ D3-branes of type $I$ is considered, 
the following phase factor is therefore introduced in the path integral 
\begin{equation}
\exp\bigg(\ii\,\sum_{I=0}^{M-1}\frac{c_I}{2\pi}\int\!d^2z_\parallel\,\mathrm{Tr}_{\mathrm{U}(n_I)}
F_I\bigg)~,
\end{equation}
leading to the following identification of the $\eta$-parameters of the surface defect:
\begin{equation}
\eta_I=\frac{c_I}{2\pi}~.
\label{etais}
\end{equation}

This completes the identification of all the parameters of the generic GW 
monodromy defect with the background values of the twisted scalars in 
the ${\mathbb Z}_M$ orbifold. We note that these formulas generalize those 
in \cite{Ashok:2020ekv} and exactly reduce to them when $M=2$. We also remark that if we write
the parameters $b_I$, $b_I^{\pm}$ and $c_I$ as sums over all twisted sectors, their relation with
the parameters of the surface defects holds also for even $M$. In this case, in fact, beside the twisted
sectors we have described at length in this paper, there is also a sector with twist $\frac{1}{2}$ whose
contribution is exactly the same as in the $M=2$ case. For this reason, therefore, we see that the
restriction we made at the beginning to restrict to odd values of $M$ does not lead to any loss of generality.

We end this section by observing that the identifications (\ref{abgare}) and (\ref{etais}), namely
\begin{equation}
\label{finalid}
\{\alpha_I, \beta_I, \gamma_I, \eta_I\} = \Big\{\!-\frac{b_I}{2\pi}, \frac{\text{Re}(b^+_{I})}{2\pi}, \frac{\text{Im}(b^+_{I})}{2\pi}, \frac{c_I}{2\pi} \Big\}~,
\end{equation}
are consistent with the behavior of the GW parameters under S-duality,
as given in \cite{Gukov:2006jk}. 
In fact, even though our world-sheet analysis has been at the orbifold fixed point, it is possible 
to blow-up the $\mathbb{Z}_M$-singularity into an ALE space and provide an interpretation to 
the twisted scalars of the orbifold theory as massless moduli in the low-energy supergravity 
(see for instance \cite{Anselmi:1993sm, Douglas:1996sw}). In such a geometric approach, the combinations $b_I$ and $c_I$, which are made of the singlets $b_{\text{s}}^{(\widehat{a})}$
and ${\mathcal C}^{(\widehat a)}$ from each twisted sector as shown in (\ref{bI}) and (\ref{cI}), 
arise by integrating, respectively, the NS/NS 2-form $B_{(2)}$ and the R/R 2-form $C_{(2)}$ of 
Type II B supergravity around the exceptional cycles $\omega_I$ of the blown-up ALE space. 
Therefore, from (\ref{finalid}) we read
\begin{equation}
\label{alphaeta}
\alpha_I = -\frac{1}{2\pi} \int_{\omega_I}B_{(2)}~, \qquad \eta_I = \frac{1}{2\pi} \int_{\omega_I}C_{(2)}~.
\end{equation}
Using the S-duality action on the 2-forms, with simple manipulations 
\cite{Ashok:2020ekv} one can show that this identification 
implies that $\alpha_I$ and $\eta_I$ indeed transform in the expected way.

Similarly, the $b^{\pm}_{I}$ parameters can be identified with the (string frame) metric moduli 
corresponding to the complex structure of the blown-up exceptional cycle $\omega_I$. As such they 
inherit the S-duality transformation properties from the (string frame) metric, which are precisely
the ones expected for the parameters $\beta_I$ and $\gamma_I$ of the GW defects.

We finally remark that when $M>2$ also the scalars $b_3^{(\widehat{a})}$ can couple to the
gauge fields, differently from what happens in the $M=2$ case \cite{Ashok:2020ekv}.
To have a uniform description for all $M$ we have therefore chosen to set 
$b^{(\widehat{a})}_{3}=0$ in each twisted sector. As we have just seen, this choice has allowed us to
identify a perturbative closed string realization of the generic GW defects that
is fully consistent with S-duality. However, our approach offers the possibility of considering
more general backgrounds with also $b^{(\widehat{a})}_{3}$ turned on, 
and it would be interesting to further investigate their meaning and 
implications for the world-volume theory on the D3-branes and their defects.

\section{Discussion} 
\label{sec:concl-m}

The present work extends the analysis of \cite{Ashok:2020ekv}, where the main ideas of our approach to a string theoretic realization of the GW surface defects were already anticipated. Here we have concentrated on the technical ingredients necessary to implement those ideas in the case of a generic half-BPS surface defect. Therefore we think it is useful to recapitulate at this point our motivations 
and the main features of our construction, and highlight some new perspectives and potential future developments.    

The study of defects in quantum field theories is an important subject from many different 
points of view.
For example, a proper understanding of conformal defects is a crucial step towards a complete classification of higher dimensional conformal field theories. In this context, much progress has 
been made in elucidating the kinematic constraints that the residual symmetry of conformal defects imposes on the observables of the theory, leading to their parametrization by some set of conformal data \cite{McAvity:1995zd,Billo:2016cpy,Gadde:2016fbj,Lauria:2017wav,Lauria:2018klo}. The kinematics is even more constrained for superconformal defects where stringent relations 
between the two-point functions of the displacement operator and the stress tensor one-point function for surface defects exist \cite{Bianchi:2019sxz}.

The general symmetry structure helps to tackle the dynamics of defects also in (super) Yang-Mills theories. The line defects corresponding to Wilson or 't Hooft lines represent a widely studied set of observables. Surface defects, whose definition is more delicate, are also extremely interesting, especially with regard to the duality properties of the theory. Groundbreaking work on conformal surface defects in gauge theories was carried out in \cite{Gukov:2006jk,Gukov:2008sn}, where half-BPS monodromy defects in $\cN=4$ super Yang-Mills theories were characterized and 
their S-duality properties clarified. Many developments followed, giving such defects an holographic realization in type IIB supergravity \cite{Gomis:2007fi, Drukker:2008wr}, extending the study to generic $\cN=2$ theories \cite{Gaiotto:2009fs} and taking advantage of $6d$ and M-theory embeddings \cite{Chacaltana:2012zy} and of localization techniques \cite{Awata:2010bz,Kanno:2011fw,Alday:2010vg, Ashok:2017odt, Gorsky:2017hro}. 

What we have done in this work is to directly realize the GW monodromy defects within perturbative 
Type II B string theory using fractional D3-branes on orbifolds. This realization was already 
suggested in \cite{Kanno:2011fw} where it was shown that the instanton contributions to the 
effective theory of surface defects are organized in terms of chain-saw quivers
and described as D-instanton corrections to a system of fractional D3-branes 
with two world-volume directions extended along the orbifold background
(see also \cite{Nawata:2014nca, Ashok:2017odt}).
Here we have taken this picture seriously and showed that such a D3-brane configuration with a partially longitudinal orbifold action is a GW defect.
It was already clear from the KT construction that the discrete data of a GW defect are 
represented by the order $M$ of the orbifold group and by the numbers $n_I$ of D3-branes 
assigned to the $I$-th irreducible representation of $\mathbb{Z}_M$. What was missing, however, 
was the identification of the non-trivial profiles of the gauge fields around the defect and 
their continuous monodromy parameters. Here we have filled this gap showing for a generic defect
how these continuous data are encoded within the D-brane configuration.

As we already pointed out in \cite{Ashok:2020ekv}, our description in terms of closed string background fields has some similarities with the holographic realization of surface defects as bubbling geometries of Type II B supergravity that asymptote to $AdS_5\times S^5$ \cite{Gomis:2007fi,Drukker:2008wr}; indeed, in that realization, like in ours, the continuous parameters 
of the defects are mapped to integrals of the NS/NS and R/R 2-forms over suitable cycles. 
Our construction, however, is based on an exactly solvable string background -- D3-branes on an orbifold -- in which explicit world-sheet computations are possible. Moreover, we are on the gauge theory side of the holographic correspondence: the branes have not dissolved into geometry and the open string degrees of freedom are explicitly present. It would be a worthwhile exercise to relate our D-branes on orbifolds to the bubbling geometries of \cite{Gomis:2007fi,Drukker:2008wr}.

For simplicity we have considered surface defects in $\cN=4$ U($N$) theories, but our analysis can be extended to cases with lower supersymmetry and/or with other gauge groups. For example, 
by introducing a mass deformation in two of the directions transverse to the D3-branes \cite{Ashok:2017odt} we can realize the so-called $\cN=2^*$ theory, or by implementing another orbifold acting purely in directions transverse to the D3-branes we can obtain a $\cN=2$ theory. Furthermore, by introducing orientifold planes we can get models and defects
with orthogonal or symplectic gauge groups. Exploring in this fashion these set-ups represents a logical line of development. 

Let us remark once more that the orbifold that realizes the GW defects
has a different behavior with respect to the orbifolds that are usually considered. In fact, as 
discussed  in Section~\ref{openspectrum}, this orbifold not only acts on the oscillators and the Chan-Paton indices of the open string states, but also on the components of their momentum transverse to the defect. This action on the momentum can therefore compensate the corresponding action on the oscillators and the Chan-Paton factors, so that no state is projected out; rather, a specific momentum dependence is imposed. Therefore, on the world-volume of the fractional D3-branes we 
find the same field content of the $\cN=4$ super Yang-Mills theory.
The exception to this pattern is represented by the open strings with no momentum transverse to the defect. Out of these states, the orbifold selects a subset of states and halves the amount of supersymmetry. Such states, which we did not investigate in the present work, represent the defect sector of the defect CFT. The bulk operators are instead represented by closed and open string vertices with non-zero momenta in the directions transverse to the defect. Mapping correlators of bulk and defect operators to ordinary string world-sheet diagrams could prove to be a useful tool in the investigation of the defect dynamics. This is another direction worth investigating.  

The perturbative string theory realization of a non-trivial sector of the gauge theory that we have described bears many analogies with the explicit derivation of the gauge instanton profiles from D3/D-instanton systems \cite{Witten:1995im,Douglas:1996uz,Green:2000ke} via the emission of open strings from disk diagrams with mixed boundary conditions \cite{Billo:2002hm}. The role of the instanton moduli is played in the construction of the surface defect by the insertion of the twisted closed string at zero momentum. The direct realization of instantons as a solvable D-brane background, besides its intrinsic interest, turned out to be very useful in evaluating instanton effects in deformed theories \cite{Billo:2004zq,Billo:2005fg,Billo:2006jm} as well as in engineering ``exotic'' instantons of purely stringy origin \cite{Blumenhagen:2006xt,Ibanez:2006da,Billo:2008pg}, possibly giving rise to effects otherwise prohibited in the effective field theory. Similarly, in the case of surface defects, it is possible that having a microscopic stringy realization might suggest some novel effects in the defect gauge theory. We hope to explore these and related issues in the future.

\vskip 1.5cm
\noindent {\large {\bf Acknowledgments}}
\vskip 0.2cm
We would like to thank Abhijit Gadde, Dileep Jatkar, Igor Pesando, Naveen Prabhakar, Madhusudhan Raman and Ashoke Sen for helpful discussions and correspondence and Renjan John for collaboration at an early stage of the project. The work of A.L. is partially supported by ``Fondi Ricerca Locale dell'Universit\`a  del Piemonte Orientale".
\vskip 1cm
\begin{appendix}
\section{Conventions}
In this appendix we list our conventions for spinors both in the 4$d$ space along 
the $\mathbb{Z}_M$ orbifold, and in the 6$d$ space transverse to it.

\subsection{Spinors in 4$d$}
\label{SO(4)}
We consider a 4$d$ space parametrized by the two complex coordinates
$z_2$ and $z_3$, related to the four real coordinates $x_m$ (with $m=3,4,5,6$) as
in (\ref{zi}). Introducing the Pauli matrices
\begin{equation}
\sigma^m=\big(\tau^1,\tau^2,\tau^3,-i\,\mathbf{1}_2\big)~,
\end{equation}
we can form the combination
\begin{equation}
X_{\alpha\dot{\beta}}=\frac{1}{\sqrt{2}} \,x_m\,({\sigma}^m)_{\alpha\dot{\beta}}=
\begin{pmatrix}
\bar{z}_3& \bar{z}_2\\
z_2& -{z}_3
\end{pmatrix}~.
\end{equation}
The $\mathrm{SO}(4)\simeq\mathrm{SU}(2)_+\times\mathrm{SU}(2)_-$ isometry group acts 
on $X$ as follows
\begin{equation}
X~~\longrightarrow~~U_+\, X\,U_-^{\dagger}
\end{equation}
where $U_\pm \in \mathrm{SU}(2)_{\pm}$.
Therefore, the two columns of X are two doublets transforming as spinors of SU$(2)_+$: 
\begin{equation}
y_\alpha=\begin{pmatrix}
\bar{z}_3\\
z_2
\end{pmatrix}
\quad\mbox{and}\quad  
w_{\alpha}=\begin{pmatrix}
\phantom{-}\bar{z}_2\\
-z_3
\end{pmatrix}
~.
\end{equation}
Raising the indices, we have
\begin{equation}
y^\alpha=y_\beta\,(\epsilon^{-1})^{\beta\alpha}=\begin{pmatrix}
-z_2\\
\phantom{-}\bar{z}_3
\end{pmatrix}
\quad\mbox{and}\quad  
w^{\alpha}=w_\beta\,(\epsilon^{-1})^{\beta\alpha}=
\begin{pmatrix}
z_3\\
\bar{z}_2
\end{pmatrix}
\end{equation}
where $\epsilon=-\ii\,\tau_2$ as in (\ref{epsilon}).
Of course the same combinations can be made with the fermionic coordinates leading
to the doublets
\begin{equation}
\begin{pmatrix}
-\Psi^2\\
\phantom{-}\overbar{\Psi}^3
\end{pmatrix}
\quad\mbox{and}\quad
\begin{pmatrix}
\Psi^3\\
\overbar{\Psi}^2
\end{pmatrix}~.
\end{equation}
These are precisely the structures that have been used in Section~\ref{closedspectrum} to write the massless vertex operators of the twisted NS/NS sectors.

\subsection{Spinors in 6$d$}
\label{twistedRspinors}

We consider a 6$d$ Euclidean space spanned by the coordinates $x_M$ with
$M\in \{1,2,7,8,9,10\}$, in order to respect the configuration of the
orbifold (\ref{orbifold}). The 6$d$ Euclidean Clifford algebra is given by
\begin{equation}
\{\Gamma_{M}, \Gamma_{N}\} = 2\delta_{MN}~,
\end{equation}
and an explicit realization of the $\Gamma$ matrices is given by: 
\begin{equation}
\begin{aligned}
\Gamma_1&=\begin{pmatrix}
0 & 0&-\ii\, \mathbb{1}_2 & 0\\
0 & 0& 0 & -\ii\,\mathbb{1}_2\\
\ii\,\mathbb{1}_2 & 0& 0 & 0\\
0 & \ii\,\mathbb{1}_2& 0 & 0\\
\end{pmatrix}~,\quad
\quad\quad
\Gamma_2=\begin{pmatrix}
0 & 0& \tau_3 & 0\\
0 & 0& 0 & -\tau_3\\
\tau_3 & 0& 0 & 0\\
0 & -\tau_3& 0 & 0\\
\end{pmatrix}~,\\
\Gamma_7&=\begin{pmatrix}
0 & 0& -\tau_2 & 0\\
0 & 0& 0 & \tau_2\\
-\tau_2 & 0& 0 & 0\\
0 & \tau_2& 0 & 0\\
\end{pmatrix}~,\quad
\qquad\quad~~~
\Gamma_8=\begin{pmatrix}
0 & 0& \tau_1 & 0\\
0 & 0& 0 & -\tau_1\\
\tau_1 & 0& 0 & 0\\
0 & -\tau_1& 0 & 0\\
\end{pmatrix}~,\\
\Gamma_9&=\begin{pmatrix}
0 & 0& 0 & -\ii\,\mathbb{1}_2\\
0 & 0& \ii\,\mathbb{1}_2 & 0\\
0 & -\ii\,\mathbb{1}_2& 0 & 0\\
\ii\,\mathbb{1}_2 & 0& 0 & 0\\
\end{pmatrix}~,\quad
\quad~~~
\Gamma_{10}=\begin{pmatrix}
0 & 0& 0 & \mathbb{1}_2\\
0 & 0& \mathbb{1}_2 & 0\\
0 & \mathbb{1}_2& 0 & 0\\
\mathbb{1}_2 & 0& 0 & 0\\
\end{pmatrix}~.\\
\end{aligned}
\end{equation}
It follows that the 6$d$ chirality matrix $\widehat{\Gamma}$ is 
\begin{equation}
\widehat{\Gamma}=\ii\,\Gamma_1\Gamma_2\Gamma_7\Gamma_8\Gamma_9\Gamma_{10}
=\begin{pmatrix}
\mathbb{1}_2 & 0 & 0 & 0\\
0& \mathbb{1}_2& 0 &0\\
0& 0&-\mathbb{1}_2& 0\\
0& 0& 0& -\mathbb{1}_2
\end{pmatrix}
~,
\label{6dchirality}
\end{equation}
which shows that, in this basis, a Dirac spinor is written as
\begin{equation}
\begin{pmatrix}
S^A\\
S^{\dot{A}}
\end{pmatrix}
\end{equation}
where $A$ and $\dot{A}$ label, respectively, the chiral and anti-chiral components. The charge conjugation matrix $C$, in this basis, is given by 
\begin{equation}
C=\begin{pmatrix}
0&0&0&\epsilon\\
0&0&\epsilon&0\\
0&-\epsilon&0&0\\
-\epsilon&0&0&0
\end{pmatrix}
\label{chargeconjugation6d}
\end{equation}
where $\epsilon=-\ii\,\tau_2$ as in (\ref{epsilon}).
The charge conjugation matrix is such that 
\begin{equation}
C\,\Gamma_M\,C^{-1}=- (\Gamma_M)^{\mathtt{t}}~.
\end{equation}
\section{$\mathbb{Z}_M$ in momentum space}
\label{sec:orbZM}

Here we briefly comment on how to define the $\mathbb{Z}_M$ orbifold action in momentum
space. Let us take the complex plane $C_{(2)}$ with coordinates $z_2$ and $\bar{z}_2$ on
which $\mathbb{Z}_M$ acts as in  (\ref{gonz23}), and define the momenta
$\kappa_2$ and $\overbar{\kappa}_2$ as in (\ref{kappai}). 
For simplicity, however, we can drop the index 2 since in this appendix 
this does not cause any ambiguity.

First of all, we observe that the orbifold action on the coordinates can be equivalently read as
an inverse action on the momenta. Consider for example the scalar product
\begin{equation}
\kappa\,\overbar{z}+\overbar{\kappa}\,z~,
\end{equation}
which, under the action of $\mathbb{Z}_M$ on the coordinates, is mapped to 
\begin{equation}
\omega^{-1}\kappa\,\overbar{z}+\omega\,\overbar{\kappa}\,z~.
\end{equation}
Clearly, this result can also be interpreted as due to the following action of $\mathbb{Z}_M$ 
on the momentum variables:
\begin{equation}
\hat{g}\,:~ (\kappa\,, \,\overbar{\kappa}) \longrightarrow (\omega^{-1}\, \kappa\,,\, \omega\, 
\overbar{\kappa})
\label{gonkappa}
\end{equation}
with the coordinates held fixed.

Then, let us consider a function in momentum space, $f(\kappa,\overbar{\kappa})$, 
and define its images under the orbifold group according to
\begin{equation}
\label{Pivsf}
\Pi_I(\kappa,\overbar{\kappa})=\frac{1}{M} \sum_{J=0}^{M-1} \omega^{-IJ} \, 
f\big(\omega^{-J}\kappa,\omega^J\,\overbar{\kappa}\big)
\end{equation}
where $I=0,\ldots,M-1$, modulo $M$.
Using (\ref{gonkappa}), it is immediate to check that
\begin{equation}
 \hat{g}\big[ \Pi_I\big]= \omega^I \, \Pi_I~,
\end{equation}
namely that $\Pi_I$ transforms in the $I$-th representation of $\mathbb{Z}_M$. Inverting
(\ref{Pivsf}), we get 
\begin{equation}
\label{fP}
f(\kappa,\overbar{\kappa})= \sum_{I=0}^{M-1} \Pi_I(\kappa,\overbar{\kappa})~.
\end{equation}
Applying these definitions to the plane wave 
$\rme^{\ii\,(\kappa\,\overbar{z}+\overbar{\kappa}\,z)}$, we get
\begin{equation}
\mathcal{E}_I=  \frac{1}{M}  \sum_{J=0}^\infty
\omega^{-IJ} \, \rme^{\ii\,(\omega^{-J}\kappa\,\overbar{z}+\omega^J\overbar{\kappa}\,z)}~,
\label{EI}
\end{equation}
with
\begin{equation}
 \hat{g}\big[ \mathcal{E}_I\big]= \omega^I \, \mathcal{E}_I~.
 \label{gonEI}
\end{equation}
These functions $\mathcal{E}_I$ have exactly the same form and properties of the functions 
introduced in Section~\ref{openspectrum} 
when we described the $\mathbb{Z}_M$-invariant open string states. In terms of them,
the plane wave can be written as
\begin{equation}
\rme^{\ii\,(\kappa\,\overbar{z}+\overbar{\kappa}\,z)}=\sum_{I=0}^{M-1}\mathcal{E}_I~.
\label{eE}
\end{equation}

Let us now consider the Fourier transform of $f$. Using (\ref{fP}) and (\ref{eE}), we have
\begin{equation}
\mathcal{FT}[f](z) =\int  \frac{d^2 \kappa}{2\pi} \,
f(\kappa,\overbar{\kappa})\, 
\rme^{\ii\,(\kappa\,\overbar{z}+\overbar{\kappa}\,z)}=
\int  \frac{d^2 \kappa}{2\pi} \,\sum_{I,J=0}^{M-1} \Pi_I(\kappa,\overbar{\kappa})\,\mathcal{E}_J~.
\end{equation}
Since the integration measure is $\mathbb{Z}_M$-invariant, only the invariant products
$\Pi_I\,\mathcal{E}_{M-I}$ survive, and thus
 \begin{equation}
\mathcal{FT}[f](z) =
\int  \frac{d^2 \kappa}{2\pi} \,\sum_{I=0}^{M-1} \Pi_I(\kappa,\overbar{\kappa})\,
\mathcal{E}_{M-I}~=~\frac{1}{M} 
\int  \frac{d^2 \kappa}{2\pi} \,
\sum_{I=0}^{M-1}  \hat{g}^I\Big[f(\kappa,\overbar{\kappa})\, 
\rme^{\ii\,(\kappa\,\overbar{z}+\overbar{\kappa}\,z)}\Big]~.
\end{equation}
This shows that the Fourier transform leads to a well-defined function in the orbifolded theory.
In particular, the Fourier transform of a function in the $I$-th irreducible reprentation of $\mathbb{Z}_M$ in momentum space is a function in configuration space that transforms
in the representation $(M-I)$, and {\emph{viceversa}}.

\end{appendix}
\endgroup

\providecommand{\href}[2]{#2}\begingroup\raggedright\endgroup


\end{document}

%% file: tadpole.pdf_tex
\begingroup%
  \makeatletter%
  \providecommand\color[2][]{%
    \errmessage{(Inkscape) Color is used for the text in Inkscape, but the package 'color.sty' is not loaded}%
    \renewcommand\color[2][]{}%
  }%
  \providecommand\transparent[1]{%
    \errmessage{(Inkscape) Transparency is used (non-zero) for the text in Inkscape, but the package 'transparent.sty' is not loaded}%
    \renewcommand\transparent[1]{}%
  }%
  \providecommand\rotatebox[2]{#2}%
  \newcommand*\fsize{\dimexpr\f@size pt\relax}%
  \newcommand*\lineheight[1]{\fontsize{\fsize}{#1\fsize}\selectfont}%
  \ifx\svgwidth\undefined%
    \setlength{\unitlength}{135bp}%
    \ifx\svgscale\undefined%
      \relax%
    \else%
      \setlength{\unitlength}{\unitlength * \real{\svgscale}}%
    \fi%
  \else%
    \setlength{\unitlength}{\svgwidth}%
  \fi%
  \global\let\svgwidth\undefined%
  \global\let\svgscale\undefined%
  \makeatother%
  \begin{picture}(1,0.59257385)%
    \lineheight{1}%
    \setlength\tabcolsep{0pt}%
    \put(0.82935067,0.31011491){\color[rgb]{0,0,0}\makebox(0,0)[lt]{\lineheight{0}\smash{\begin{tabular}[t]{l}
    \end{tabular}}}}%
    \put(0,0){\includegraphics[width=\unitlength,page=1]{tadpole.pdf}}%
    \put(0.06179471,0.37063218){\color[rgb]{0,0,0}\makebox(0,0)[lt]{\lineheight{1.25}\smash{\begin{tabular}[t]{l}\textbf{$\cV_{\mathrm{open}}$}\end{tabular}}}}%
    \put(0.46513473,0.18771536){\color[rgb]{0,0,0}\makebox(0,0)[lt]{\lineheight{1.25}\smash{\begin{tabular}[t]{l}{\small\textbf{$V_{b_{\alpha\beta}^{(\widehat{a})}}$}}\end{tabular}}}}%
    \put(0.73370991,0.52739252){\color[rgb]{0,0,0}\makebox(0,0)[lt]{\lineheight{1.25}\smash{\begin{tabular}[t]{l}\textbf{$\mathrm{D3}_I$}\end{tabular}}}}%
    \put(0,0){\includegraphics[width=\unitlength,page=2]{tadpole.pdf}}%
    \put(0.0464934,0.19462331){\color[rgb]{0,0,0}\makebox(0,0)[lt]{\lineheight{1.25}\smash{\begin{tabular}[t]{l}\textbf{$\vec{k}_\perp$}\end{tabular}}}}%
  \end{picture}%
\endgroup%

%% file: surf_op_ZM_a.bbl
\begin{thebibliography}{10}

\bibitem{Ashok:2020ekv}
S.~Ashok, M.~Billo, M.~Frau, A.~Lerda, and S.~Mahato, \emph{{Surface Defects
  from Fractional Branes -- I}}, \href{http://arxiv.org/abs/2005.02050}{{\tt
  arXiv:2005.02050 [hep-th]}}.

\bibitem{Gukov:2006jk}
S.~Gukov and E.~Witten, \emph{{Gauge Theory, Ramification, And The Geometric
  Langlands Program}},
\href{http://arxiv.org/abs/hep-th/0612073}{{\tt arXiv:hep-th/0612073
  [hep-th]}}.

\bibitem{Gukov:2008sn}
S.~Gukov and E.~Witten, \emph{{Rigid Surface Operators}},
  \href{http://dx.doi.org/10.4310/ATMP.2010.v14.n1.a3}{Adv. Theor. Math. Phys.
  {\bf 14} (2010) no.~1, 87--178},
\href{http://arxiv.org/abs/0804.1561}{{\tt arXiv:0804.1561 [hep-th]}}.

\bibitem{Kanno:2011fw}
H.~Kanno and Y.~Tachikawa, \emph{{Instanton counting with a surface operator
  and the chain-saw quiver}},
  \href{http://dx.doi.org/10.1007/JHEP06(2011)119}{JHEP {\bf 06} (2011)  119},
\href{http://arxiv.org/abs/1105.0357}{{\tt arXiv:1105.0357 [hep-th]}}.

\bibitem{Friedan:1985ge}
D.~Friedan, E.~J. Martinec, and S.~H. Shenker, \emph{{Conformal Invariance,
  Supersymmetry and String Theory}},
Nucl. Phys. {\bf B271} (1986)  93.

\bibitem{Kostelecky:1986xg}
V.~A. Kostelecky, O.~Lechtenfeld, W.~Lerche, S.~Samuel, and S.~Watamura,
  \emph{{Conformal Techniques, Bosonization and Tree Level String Amplitudes}},
\href{http://dx.doi.org/10.1016/0550-3213(87)90213-6}{Nucl. Phys. {\bf B288}
  (1987)  173--232}.

\bibitem{DiVecchia:1999mal}
P.~Di~Vecchia and A.~Liccardo, \emph{{D Branes in String Theory. 1}},
  \href{http://dx.doi.org/10.1007/978-94-011-4303-5_1}{NATO Sci. Ser. C {\bf
  556} (2000)  1--60},
\href{http://arxiv.org/abs/hep-th/9912161}{{\tt arXiv:hep-th/9912161
  [hep-th]}}.

\bibitem{DiVecchia:1999fje}
P.~Di~Vecchia and A.~Liccardo, \emph{{D branes in String Theory. 2.}}, in {\em
  {YITP Workshop on Developments in Superstring and M Theory Kyoto, Japan,
  October 27-29, 1999}}, pp.~7--48.
\newblock 1999.
\newblock
\href{http://arxiv.org/abs/hep-th/9912275}{{\tt arXiv:hep-th/9912275
  [hep-th]}}.
\newblock

\bibitem{Douglas:1996sw}
M.~R. Douglas and G.~W. Moore, \emph{{D-branes, Quivers, and ALE Instantons}},
\href{http://arxiv.org/abs/hep-th/9603167}{{\tt arXiv:hep-th/9603167}}.

\bibitem{Billo:2000yb}
M.~Billo, B.~Craps, and F.~Roose, \emph{{Orbifold boundary states from Cardy's
  condition}}, JHEP {\bf 01} (2001)  038,
\href{http://arxiv.org/abs/hep-th/0011060}{{\tt arXiv:hep-th/0011060}}.

\bibitem{Billo:2001vg}
M.~Billo, L.~Gallot, and A.~Liccardo, \emph{{Classical geometry and gauge duals
  for fractional branes on ALE orbifolds}},
  \href{http://dx.doi.org/10.1016/S0550-3213(01)00399-6}{Nucl. Phys. {\bf B614}
  (2001)  254--278},
\href{http://arxiv.org/abs/hep-th/0105258}{{\tt arXiv:hep-th/0105258}}.

\bibitem{Bertolini:2001gq}
M.~Bertolini, P.~Di~Vecchia, and R.~Marotta, \emph{{N=2 four-dimensional gauge
  theories from fractional branes}},
\href{http://arxiv.org/abs/hep-th/0112195}{{\tt arXiv:hep-th/0112195
  [hep-th]}}.

\bibitem{Bertolini:2005qh}
M.~Bertolini, M.~Billo, A.~Lerda, J.~F. Morales, and R.~Russo, \emph{{Brane
  world effective actions for D-branes with fluxes}},
  \href{http://dx.doi.org/10.1016/j.nuclphysb.2006.02.044}{Nucl. Phys. {\bf
  B743} (2006)  1--40},
\href{http://arxiv.org/abs/hep-th/0512067}{{\tt arXiv:hep-th/0512067}}.

\bibitem{Dixon:1986qv}
L.~J. Dixon, D.~Friedan, E.~J. Martinec, and S.~H. Shenker, \emph{{The
  Conformal Field Theory of Orbifolds}},
\href{http://dx.doi.org/10.1016/0550-3213(87)90676-6}{Nucl. Phys. {\bf B282}
  (1987)  13--73}.

\bibitem{Billo:1998vr}
M.~Billo, P.~Di~Vecchia, M.~Frau, A.~Lerda, I.~Pesando, R.~Russo, and
  S.~Sciuto, \emph{{Microscopic string analysis of the D0-D8 brane system and
  dual R-R states}},
  \href{http://dx.doi.org/10.1016/S0550-3213(98)00296-X}{Nucl. Phys. {\bf B526}
  (1998)  199--228},
\href{http://arxiv.org/abs/hep-th/9802088}{{\tt arXiv:hep-th/9802088}}.

\bibitem{Cardy:1989ir}
J.~L. Cardy, \emph{{Boundary Conditions, Fusion Rules and the Verlinde
  Formula}},
\href{http://dx.doi.org/10.1016/0550-3213(89)90521-X}{Nucl. Phys. {\bf B324}
  (1989)  581}.

\bibitem{Hashimoto:1996bf}
A.~Hashimoto and I.~R. Klebanov, \emph{{Scattering of strings from D-branes}},
  \href{http://dx.doi.org/10.1016/S0920-5632(97)00074-1}{Nucl. Phys. B Proc.
  Suppl. {\bf 55} (1997)  118--133},
  \href{http://arxiv.org/abs/hep-th/9611214}{{\tt arXiv:hep-th/9611214}}.

\bibitem{Anselmi:1993sm}
D.~Anselmi, M.~Billo, P.~Fre, L.~Girardello, and A.~Zaffaroni, \emph{{ALE
  manifolds and conformal field theories}},
  \href{http://dx.doi.org/10.1142/S0217751X94001199}{Int. J. Mod. Phys. {\bf
  A9} (1994)  3007--3058},
\href{http://arxiv.org/abs/hep-th/9304135}{{\tt arXiv:hep-th/9304135
  [hep-th]}}.

\bibitem{McAvity:1995zd}
D.~M. McAvity and H.~Osborn, \emph{{Conformal field theories near a boundary in
  general dimensions}},
  \href{http://dx.doi.org/10.1016/0550-3213(95)00476-9}{Nucl. Phys. {\bf B455}
  (1995)  522--576},
\href{http://arxiv.org/abs/cond-mat/9505127}{{\tt arXiv:cond-mat/9505127
  [cond-mat]}}.

\bibitem{Billo:2016cpy}
M.~Billo, V.~Goncalves, E.~Lauria, and M.~Meineri, \emph{{Defects in conformal
  field theory}}, \href{http://dx.doi.org/10.1007/JHEP04(2016)091}{JHEP {\bf
  04} (2016)  091},
\href{http://arxiv.org/abs/1601.02883}{{\tt arXiv:1601.02883 [hep-th]}}.

\bibitem{Gadde:2016fbj}
A.~Gadde, \emph{{Conformal constraints on defects}},
  \href{http://dx.doi.org/10.1007/JHEP01(2020)038}{JHEP {\bf 01} (2020)  038},
  \href{http://arxiv.org/abs/1602.06354}{{\tt arXiv:1602.06354 [hep-th]}}.

\bibitem{Lauria:2017wav}
E.~Lauria, M.~Meineri, and E.~Trevisani, \emph{{Radial coordinates for defect
  CFTs}}, \href{http://dx.doi.org/10.1007/JHEP11(2018)148}{JHEP {\bf 11} (2018)
   148}, \href{http://arxiv.org/abs/1712.07668}{{\tt arXiv:1712.07668
  [hep-th]}}.

\bibitem{Lauria:2018klo}
E.~Lauria, M.~Meineri, and E.~Trevisani, \emph{{Spinning operators and defects
  in conformal field theory}},
  \href{http://dx.doi.org/10.1007/JHEP08(2019)066}{JHEP {\bf 08} (2019)  066},
  \href{http://arxiv.org/abs/1807.02522}{{\tt arXiv:1807.02522 [hep-th]}}.

\bibitem{Bianchi:2019sxz}
L.~Bianchi and M.~Lemos, \emph{{Superconformal surfaces in four dimensions}},
  \href{http://arxiv.org/abs/1911.05082}{{\tt arXiv:1911.05082 [hep-th]}}.

\bibitem{Gomis:2007fi}
J.~Gomis and S.~Matsuura, \emph{Bubbling surface operators and S-duality},
  \href{http://dx.doi.org/10.1088/1126-6708/2007/06/025}{JHEP {\bf 06} (2007)
  025}, \href{http://arxiv.org/abs/0704.1657}{{\tt arXiv:0704.1657 [hep-th]}}.

\bibitem{Drukker:2008wr}
N.~Drukker, J.~Gomis, and S.~Matsuura, \emph{Probing N=4 SYM With Surface
  Operators}, \href{http://dx.doi.org/10.1088/1126-6708/2008/10/048}{JHEP {\bf
  10} (2008)  048}, \href{http://arxiv.org/abs/0805.4199}{{\tt arXiv:0805.4199
  [hep-th]}}.

\bibitem{Gaiotto:2009fs}
D.~Gaiotto, \emph{{Surface Operators in N = 2 4d Gauge Theories}},
  \href{http://dx.doi.org/10.1007/JHEP11(2012)090}{JHEP {\bf 11} (2012)  090},
\href{http://arxiv.org/abs/0911.1316}{{\tt arXiv:0911.1316 [hep-th]}}.

\bibitem{Chacaltana:2012zy}
O.~Chacaltana, J.~Distler, and Y.~Tachikawa, \emph{{Nilpotent orbits and
  codimension-two defects of 6d N=(2,0) theories}},
  \href{http://dx.doi.org/10.1142/S0217751X1340006X}{Int. J. Mod. Phys. {\bf
  A28} (2013)  1340006},
\href{http://arxiv.org/abs/1203.2930}{{\tt arXiv:1203.2930 [hep-th]}}.

\bibitem{Awata:2010bz}
H.~Awata, H.~Fuji, H.~Kanno, M.~Manabe, and Y.~Yamada, \emph{{Localization with
  a Surface Operator, Irregular Conformal Blocks and Open Topological String}},
  \href{http://dx.doi.org/10.4310/ATMP.2012.v16.n3.a1}{Adv. Theor. Math. Phys.
  {\bf 16} (2012) no.~3, 725--804},
\href{http://arxiv.org/abs/1008.0574}{{\tt arXiv:1008.0574 [hep-th]}}.

\bibitem{Alday:2010vg}
L.~F. Alday and Y.~Tachikawa, \emph{{Affine SL(2) conformal blocks from 4d
  gauge theories}}, \href{http://dx.doi.org/10.1007/s11005-010-0422-4}{Lett.
  Math. Phys. {\bf 94} (2010)  87--114},
\href{http://arxiv.org/abs/1005.4469}{{\tt arXiv:1005.4469 [hep-th]}}.

\bibitem{Ashok:2017odt}
S.~K. Ashok, M.~Billo, E.~Dell'Aquila, M.~Frau, R.~R. John, and A.~Lerda,
  \emph{{Modular and duality properties of surface operators in N=2* gauge
  theories}}, \href{http://dx.doi.org/10.1007/JHEP07(2017)068}{JHEP {\bf 07}
  (2017)  068},
\href{http://arxiv.org/abs/1702.02833}{{\tt arXiv:1702.02833 [hep-th]}}.

\bibitem{Gorsky:2017hro}
A.~Gorsky, B.~Le~Floch, A.~Milekhin, and N.~Sopenko, \emph{{Surface defects and
  instanton/vortex interaction}},
  \href{http://dx.doi.org/10.1016/j.nuclphysb.2017.04.010}{Nucl. Phys. {\bf
  B920} (2017)  122--156},
\href{http://arxiv.org/abs/1702.03330}{{\tt arXiv:1702.03330 [hep-th]}}.

\bibitem{Nawata:2014nca}
S.~Nawata, \emph{{Givental J-functions, Quantum integrable systems, AGT
  relation with surface operator}},
  \href{http://dx.doi.org/10.4310/ATMP.2015.v19.n6.a4}{Adv. Theor. Math. Phys.
  {\bf 19} (2015)  1277--1338},
\href{http://arxiv.org/abs/1408.4132}{{\tt arXiv:1408.4132 [hep-th]}}.

\bibitem{Witten:1995im}
E.~Witten, \emph{{Bound states of strings and p-branes}},
  \href{http://dx.doi.org/10.1016/0550-3213(95)00610-9}{Nucl. Phys. B {\bf 460}
  (1996)  335--350}, \href{http://arxiv.org/abs/hep-th/9510135}{{\tt
  arXiv:hep-th/9510135}}.

\bibitem{Douglas:1996uz}
M.~R. Douglas, \emph{{Gauge fields and D-branes}},
  \href{http://dx.doi.org/10.1016/S0393-0440(97)00024-7}{J. Geom. Phys. {\bf
  28} (1998)  255--262}, \href{http://arxiv.org/abs/hep-th/9604198}{{\tt
  arXiv:hep-th/9604198}}.

\bibitem{Green:2000ke}
M.~B. Green and M.~Gutperle, \emph{{D-instanton induced interactions on a
  D3-brane}}, JHEP {\bf 02} (2000)  014,
\href{http://arxiv.org/abs/hep-th/0002011}{{\tt arXiv:hep-th/0002011}}.

\bibitem{Billo:2002hm}
M.~Billo, M.~Frau, I.~Pesando, F.~Fucito, A.~Lerda, and A.~Liccardo,
  \emph{{Classical gauge instantons from open strings}}, JHEP {\bf 02} (2003)
  045,
\href{http://arxiv.org/abs/hep-th/0211250}{{\tt arXiv:hep-th/0211250}}.

\bibitem{Billo:2004zq}
M.~Billo, M.~Frau, I.~Pesando, and A.~Lerda, \emph{{N = 1/2 gauge theory and
  its instanton moduli space from open strings in R-R background}}, JHEP {\bf
  05} (2004)  023,
\href{http://arxiv.org/abs/hep-th/0402160}{{\tt arXiv:hep-th/0402160}}.

\bibitem{Billo:2005fg}
M.~Billo, M.~Frau, S.~Sciuto, G.~Vallone, and A.~Lerda, \emph{{Non-commutative
  (D)-instantons}}, JHEP {\bf 05} (2006)  069,
\href{http://arxiv.org/abs/hep-th/0511036}{{\tt arXiv:hep-th/0511036}}.

\bibitem{Billo:2006jm}
M.~Billo, M.~Frau, F.~Fucito, and A.~Lerda, \emph{{Instanton calculus in R-R
  background and the topological string}}, JHEP {\bf 11} (2006)  012,
\href{http://arxiv.org/abs/hep-th/0606013}{{\tt arXiv:hep-th/0606013}}.

\bibitem{Blumenhagen:2006xt}
R.~Blumenhagen, M.~Cvetic, and T.~Weigand, \emph{{Spacetime instanton
  corrections in 4D string vacua - the seesaw mechanism for D-brane models}},
  \href{http://dx.doi.org/10.1016/j.nuclphysb.2007.02.016}{Nucl. Phys. {\bf
  B771} (2007)  113--142},
\href{http://arxiv.org/abs/hep-th/0609191}{{\tt arXiv:hep-th/0609191}}.

\bibitem{Ibanez:2006da}
L.~E. Ibanez and A.~M. Uranga, \emph{{Neutrino Majorana masses from string
  theory instanton effects}}, JHEP {\bf 03} (2007)  052,
\href{http://arxiv.org/abs/hep-th/0609213}{{\tt arXiv:hep-th/0609213}}.

\bibitem{Billo:2008pg}
M.~Billo, L.~Ferro, M.~Frau, F.~Fucito, A.~Lerda, and J.~F. Morales,
  \emph{{Non-perturbative effective interactions from fluxes}},
  \href{http://dx.doi.org/10.1088/1126-6708/2008/12/102}{JHEP {\bf 12} (2008)
  102}, \href{http://arxiv.org/abs/0807.4098}{{\tt arXiv:0807.4098 [hep-th]}}.

\end{thebibliography}
